\documentclass[pdflatex,sn-mathphys-num]{sn-jnl}

\usepackage{graphicx}%
\usepackage{multirow}%
\usepackage{amsmath,amssymb,amsfonts}%
\usepackage{amsthm}%
\usepackage{mathrsfs}%
\usepackage[title]{appendix}%
\usepackage{xcolor}%
\usepackage{textcomp}%
\usepackage{manyfoot}%
\usepackage{booktabs}%
\usepackage{algorithm}%
\usepackage{algorithmicx}%
\usepackage{algpseudocode}%
\usepackage{listings}%
\usepackage{bbding}
\usepackage{array}

\usepackage{rotating}

\usepackage{bbm}
\usepackage{color,xcolor}
\usepackage{colortbl}
\definecolor{mygray}{gray}{.9}
\usepackage{caption}

\theoremstyle{thmstyleone}%
\theoremstyle{thmstyletwo}%

\theoremstyle{thmstylethree}%

\begin{document}


\title[Article Title]{FreeTumor: Large-Scale Generative Tumor Synthesis in Computed Tomography Images for Improving Tumor Recognition}


\author[1]{\fnm{Linshan} \sur{Wu}}

\author[1]{\fnm{Jiaxin} \sur{Zhuang}}

\author[2]{\fnm{Yanning} \sur{Zhou}}

\author[1]{\fnm{Sunan} \sur{He}}

\author[1]{\fnm{Jiabo} \sur{Ma}}

\author[1,3]{\fnm{Luyang} \sur{Luo}}

\author[4]{\fnm{Xi} \sur{Wang}}

\author[1]{\fnm{Xuefeng} \sur{Ni}}

\author[5]{\fnm{Xiaoling} \sur{Zhong}}

\author[5]{\fnm{Mingxiang} \sur{Wu}}

\author[6]{\fnm{Yinghua} \sur{Zhao}}

\author[7]{\fnm{Xiaohui} \sur{Duan}}

\author[8]{\fnm{Varut} \sur{Vardhanabhuti}}

\author[3]{\fnm{Pranav} \sur{Rajpurkar}}

\author*[1,9,10,11,12]{\fnm{Hao} \sur{Chen}}\email{jhc@cse.ust.hk}

\affil[1]{\orgdiv{Department of Computer Science and Engineering}, \orgname{The Hong Kong University of
Science and Technology}, \orgaddress{\state{Hong Kong}, \country{China}}}

\affil[2]{\orgdiv{Tencent AI Lab}, \orgaddress{\state{Shenzhen}, \country{China}}}

\affil[3]{\orgdiv{Department of Biomedical Informatics}, \orgname{Harvard University}, \orgaddress{\state{Boston}, \country{USA}}}

\affil[4]{\orgdiv{Department of Computer Science and Engineering}, \orgname{The Chinese University of
Hong Kong}, \orgaddress{\state{Hong Kong}, \country{China}}}

\affil[5]{\orgdiv{Department of Radiology}, \orgname{Shenzhen People’s Hospital}, \orgaddress{\state{Shenzhen}, \country{China}}}

\affil[6]{\orgdiv{Department of Radiology}, \orgname{The
Third Affiliated Hospital of Southern Medical University}, \orgaddress{\state{Guangzhou}, \country{China}}}

\affil[7]{\orgdiv{Department of Radiology}, \orgname{Sun Yat-Sen Memorial Hospital, Sun Yat-Sen University}, 
\orgaddress{\state{Guangzhou}, \country{China}}}

\affil[8]{\orgdiv{Department of Diagnostic Radiology}, \orgname{Li Ka Shing Faculty of Medicine, The University of Hong Kong}, \orgaddress{\state{Hong Kong}, \country{China}}}

\affil[9]{\orgdiv{Department of Chemical and Biological Engineering}, \orgname{The Hong Kong University of Science and Technology}, \orgaddress{\state{Hong Kong}, \country{China}}}

\affil[10]{\orgdiv{Division of Life Science}, \orgname{The Hong Kong University of Science and Technology}, \orgaddress{\state{Hong Kong}, \country{China}}}

\affil[11]{\orgdiv{State Key Laboratory of Molecular Neuroscience}, \orgname{The Hong Kong University of Science and Technology}, \orgaddress{\state{Hong Kong}, \country{China}}}

\affil[12]{\orgdiv{Shenzhen-Hong Kong Collaborative Innovation Research Institute}, \orgname{The Hong Kong University of Science and Technology}, \orgaddress{\state{Shenzhen}, \country{China}}}



\abstract{Tumor is a leading cause of death worldwide, with an estimated 10 million deaths attributed to tumor-related diseases every year.
AI-driven tumor recognition unlocks new possibilities for more precise and intelligent tumor screening and diagnosis. 
However, the progress is heavily hampered by the scarcity of annotated datasets, which demands extensive annotation efforts by radiologists. 
To tackle this challenge, we introduce \textbf{FreeTumor}, an innovative Generative AI (GAI) framework to enable large-scale tumor synthesis for mitigating data scarcity.
Specifically, FreeTumor effectively leverages a combination of limited labeled data and large-scale unlabeled data for tumor synthesis training.
Unleashing the power of large-scale data, FreeTumor is capable of synthesizing a large number of realistic tumors on images for augmenting training datasets. 
To this end, we create the largest training dataset for tumor synthesis and recognition by curating 161,310 publicly available Computed Tomography (CT) volumes from 33 sources, with only 2.3\% containing annotated tumors. 
To validate the fidelity of synthetic tumors, we engaged 13 board-certified radiologists in a Visual Turing Test to discern between synthetic and real tumors. 
Rigorous clinician evaluation validates the high quality of our synthetic tumors, as they achieved only 51.1\% sensitivity and 60.8\% accuracy in distinguishing our synthetic tumors from real ones.
Through high-quality tumor synthesis, FreeTumor scales up the recognition training datasets by over 40 times, showcasing a notable superiority over state-of-the-art AI methods including various synthesis methods and foundation models. On average, FreeTumor improves the segmentation Dice scores by 6.7\% and early tumor detection sensitivity by 16.4\%. 
These findings indicate promising prospects of FreeTumor in clinical applications, potentially advancing tumor treatments and improving the survival rates of patients.
}





\maketitle

\section{Introduction}\label{sec1}



Tumors contribute significantly to the global burden of disease, accounting for an estimated 10 million deaths annually according to the findings of the World Health Organization~\cite{bray2024global}. With the rapid advancements of deep learning~\cite{lecun2015deep,jumper2021highly,resnet,imagenet,krizhevsky2012imagenet}, AI-driven tumor recognition~\cite{biomedparse,nnUNet,medsam,msd,peiris2023uncertainty,wang2021annotation,cao2023large,sun2024foundation,avram2024accurate} has received increasing attention in clinical applications. However, existing tumor recognition methods heavily rely on annotated tumor datasets for training~\cite{biomedparse,nnUNet,medsam,cao2023large,wang2024self}, demanding substantial medical expertise and dedicated efforts for data collection and annotation. Suffering from the data-hungry nature of AI methods and the extensive annotation burden, the limited scale of tumor datasets significantly poses a substantial obstacle to the advancement of AI-driven tumor recognition.

To address this challenge, data augmentation with synthetic data has emerged as a potential solution. Recently, Generative AI (GAI)~\cite{zhu2017unpaired,isola2017image,diffusion,ldm,zhang2023adding,wu2024freetumor} has witnessed rapid development, which can generate large-scale realistic images, presenting a potential solution to mitigate the scarcity of annotated datasets~\cite{chen2021synthetic}. Specifically, synthetic data can increase the scale and diversity of training datasets, significantly boosting the robustness and generalization of AI models~\cite{depthanythingV2,oasis,fan2024scaling,freemask,wu2023querying,zhong2020random}. 
GAI has also attracted increasing attention in medical research~\cite{wang2024self,bluethgen2024vision,peng2023ai,jo2023promise,tudosiu2024realistic,degrave2023auditing,ktena2024generative,gao2023synthetic,schafer2024overcoming,carrillo2024generation}, demonstrating that GAI can synthesize high-quality medical images and consequently enhancing medical image understanding. Although encouraging results have been demonstrated, previous works largely ignored the importance of tumor synthesis, leading to limited improvements in downstream tumor recognition tasks~\cite{nnUNet,voco}.

In this study, we explore GAI to synthesize high-quality tumors on images, aiming to mitigate the scarcity of annotated tumor datasets. Early attempts~\cite{Syntumor,lyu2022pseudo,yao2021label,wang2022anomaly,wyatt2022anoddpm} utilized handcrafted image processing techniques to synthesize tumors on images. However, these handcrafted methods require complex designs from radiologists and the synthetic tumors still differ significantly from real tumors, thus failing to improve the downstream performance effectively. Recently, diffusion models, especially conditioned diffusion models~\cite{diffusion,ldm,zhang2023adding,croitoru2023diffusion,ddpm,songdenoising} have received increasing attention in recent advances of GAI. 
Although with promising achievements, these conditioned diffusion models heavily rely on the guidance of conditioning information, \emph{e.g.}, text or mask annotations. 
Thus, when applying conditioned diffusion models to tumor synthesis~\cite{Difftumor}, the synthesis training is still limited by the scale of annotated tumor datasets and falls short in leveraging large-scale data. Constrained by the scale of training datasets, conditioned diffusion models may encounter challenges in effectively generalizing to extensive unseen datasets from various sources, particularly when faced with a wide range of diverse medical image characteristics such as varying intensity levels, spacing patterns, and resolutions.

Our goal is to unleash the power of large-scale unlabeled data via high-quality tumor synthesis, aiming to augment training datasets and fortify the foundations of tumor recognition. The primary challenges include: (1) effectively leveraging large-scale unlabeled data for tumor synthesis training and (2) synthesizing realistic tumors for segmentation training.
Confronted with the challenge of conditioned diffusion models lacking the ability to leverage large-scale unlabeled data, our focus shifts towards the exploration of adversarial-training methods, \emph{i.e.}, Generative Adversarial Networks (GAN)~\cite{zhu2017unpaired,isola2017image,SPADE,oasis}. GAN-based methods involve training a generator for data generation and a discriminator for distinguishing between real and generated data, which excels in leveraging unpaired data for synthesis training. Specifically, we investigate adversarial-training methods to tackle the two aforementioned challenges: (1) The adversarial-training methods for unpaired data facilitate the integration of large-scale unlabeled data into tumor synthesis training, \emph{i.e.}, train a generator to synthesize tumors on unlabeled images and discriminate them with a discriminator (real or synthetic tumors). (2) The incorporated discriminator further enables us to discard the low-quality synthetic tumors, \emph{i.e.}, synthetic tumors failing to pass the discriminator will be discarded, thus facilitating quality control of synthetic tumors for boosting subsequent segmentation training.

\begin{figure*}
	\centering
	\includegraphics[width=1\linewidth]{./fig/overview.pdf}
\end{figure*}
\clearpage
\begin{figure*}
	\centering
	\caption{\textbf{Overview of the study}. \textbf{a.} We explore tumor synthesis and segmentation on five types of tumors/lesions, \emph{i.e.}, liver tumors, pancreas tumors, kidney tumors, lung tumors, and COVID-19. \textbf{b.} The rapid advancements in medical imaging have enabled the collection of large-scale Computed Tomography (CT) data. However, annotated tumor datasets are scarce due to the extensive annotation burden. \textbf{c.} We curated 161,310 CT volumes from 33 public sources to enable large-scale tumor synthesis and recognition, with merely $2.3\%$ of them comprising annotated tumors. \textbf{d.} FreeTumor consists of two stages: synthesis training and segmentation training. In Stage 1, FreeTumor effectively unleashes the power of large-scale unlabeled data for tumor synthesis training. In Stage 2, FreeTumor synthesizes high-quality tumors on healthy organs, facilitating the integration of large-scale unlabeled data in tumor segmentation training. \textbf{e.} Clinical evaluation of synthetic tumors. We invited 13 board-certified radiologists to a Visual Turing Test to discern between synthetic and real tumors. Rigorous clinician evaluation validates the high quality of our synthetic tumors. \textbf{f.} Extensive segmentation results on 12 public datasets showcase the superiority of FreeTumor. Specifically, FreeTumor adopts SwinUNETR~\cite{swin} as the segmentation model and employs tumor synthesis for augmenting segmentation datasets. With large-scale synthetic tumors for training, FreeTumor surpasses the baseline SwinUNETR~\cite{swin} by significant margins, achieving $10.6\%$, $5.5\%$, $3.8\%$, $6.1\%$, and $7.9\%$ Dice score improvements for five types of tumors/lesions, respectively. \textbf{g.} Early tumor detection results. With tumor synthesis, FreeTumor yields average $+16.4\%$ sensitivity improvements.}
	\label{fig_overview}
\end{figure*}

To this end, we introduce FreeTumor, \textbf{the first GAI framework tailored for large-scale tumor synthesis and segmentation training.}
FreeTumor can synthesize high-quality tumors on healthy organs without the requirement of extra annotations from radiologists. This innovation
facilitates the integration of large-scale unlabeled data into segmentation training.
As illustrated in \textbf{Figure~\ref{fig_overview}~(d)}, FreeTumor operates through two pivotal stages: synthesis training and segmentation training. In Stage 1, FreeTumor effectively leverages a combination of limited labeled data and large-scale unlabeled data for adversarial-based tumor synthesis training. Subsequently, in Stage 2, FreeTumor is employed to synthesize tumors on healthy organs for segmentation training. Simultaneously, FreeTumor incorporates a discriminator to discard low-quality synthetic tumors, enabling automatic quality control of large-scale synthetic tumors. 
By integrating large-scale datasets from diverse sources for synthesis training, FreeTumor significantly improves the quantity, quality, and diversity of tumors for training, enhancing the robustness of tumor recognition.






To evaluate the effectiveness of FreeTumor in leveraging large-scale data, we create the largest training dataset for tumor synthesis and recognition by curating 161,310 publicly available CT volumes from different medical centers, with only $2.3\%$ of them comprising annotated tumors. 
We evaluate the effectiveness of FreeTumor across four types of tumors, \emph{i.e.}, liver tumors, pancreas tumors, kidney tumors, and lung tumors. 
FreeTumor is versatile and can also be applied for COVID-19 lesions.
To validate the fidelity of synthetic tumors, we engage 13 board-certified radiologists in a Visual Turing Test to discern between synthetic and real tumors. 
Rigorous clinician evaluation validates the high quality of our synthesis results, as they achieved only $51.1\%$ sensitivity and $60.8\%$ accuracy in distinguishing our synthetic tumors from real ones. 
Extensive experiments on 12 public datasets highlight the superiority of FreeTumor. Augmenting the training datasets by over 40 times, FreeTumor clearly surpasses state-of-the-art AI methods~\cite{UNET,transunet,unetr,swin,nnUNet,Syntumor,Difftumor,MAE,SwinSSL,voco-v1}, including various synthesis methods and foundation models. 
Furthermore, the synthesis of small tumors can enhance the performance of early tumor detection, substantially aiding the timely treatment of patients. These findings underscore the promising potential of FreeTumor in improving tumor recognition within clinical practice.


\section{Results}\label{sec2}

\textbf{Datasets}. The rapid advancements in medical imaging have enabled the collection of large-scale CT data. However, few previous works have considered harnessing the untapped potential of large-scale unlabeled CT data for tumor recognition~\cite{voco}. As shown in \textbf{Figure~\ref{fig_overview}~(c)}, we curate the existing largest training dataset for tumor synthesis and recognition, encompassing 161,310 publicly available CT volumes from 33 different sources. It is worth noting that only $2.3\%$ of them (3,696 volumes) contain annotated tumors. The pre-processing details of datasets are presented in \textbf{Section~\ref{sec_method_implementation}}. Details of datasets are presented in \textbf{Extended Data Table~\ref{table_dataset}}.

\textbf{Clinician evaluation of synthetic tumors}. It has been a common practice to utilize fidelity metrics like Fréchet Inception Distance (FID)~\cite{fid} to measure the quality of natural image synthesis in GAI models~\cite{zhu2017unpaired,isola2017image,diffusion,ldm,zhang2023adding}, where lower FIDs reflects higher synthesis quality.
We first evaluate the FID results of our synthetic tumors, detailed FID results are presented in \textbf{Extended Data Table~\ref{table_fid} and Figure~\ref{fig_FID}}. We observe that our proposed FreeTumor can achieve lower FID compared with two previous tumor synthesis methods~\cite{Syntumor,Difftumor}.
However, we have noted limitations in the effectiveness of FID~\cite{fid} in reflecting tumor synthesis quality. Specifically, many synthetic tumors, despite with low FIDs, still present with unrealistic characteristics in the views of radiologists. The inherent challenge lies in the fact that tumor regions predominantly exhibit small sizes with abnormal intensities, rendering conventional fidelity metrics unreliable~\cite{wang2024self,Syntumor,Difftumor}. 
Clinician evaluation serves as a more convincing standard for validating the quality of tumor synthesis.
To this end, we invited 13 board-certified radiologists to evaluate the quality of synthetic tumors.

\textbf{Evaluation of tumor segmentation and detection}. Tumor segmentation~\cite{nnUNet,biomedparse,medsam} aims to precisely segment target tumors by capturing their positions, sizes, and shapes. In contrast, tumor detection~\cite{cao2023large,Difftumor,biomedparse} focuses on identifying the presence and location of tumors, without the need to outline their precise shapes and sizes. Following previous methods~\cite{cao2023large,fitzgerald2022future,singhi2019early,pereira2020early,bassi2025radgpt,Difftumor}, tumor detection is also achieved by the tumor segmentation models, where detected tumors are identified when segmentation predictions overlap with ground truth labels. For the evaluation of early tumor detection, we present the detection results of small tumors (diameter $<$ 2cm)~\cite{bassi2025radgpt,cao2023large}. The diameter measurement follows the standard of the World Health Organization (WHO)~\cite{miller1981reporting,bassi2025radgpt}.



We evaluate the effectiveness of FreeTumor across four types of tumors, \emph{i.e.}, liver tumors, pancreas tumors, kidney tumors, and lung tumors. FreeTumor is versatile and can
also be applied for COVID-19 lesions. We assess the performance of these five types of tumors/lesions due to the availability of public annotated datasets for evaluation. 
As shown in \textbf{Figure~\ref{fig_overview}~(e)}, 12 public datasets are used to evaluate the performances of tumor segmentation and detection, including: (1) Liver tumors: LiTS~\cite{LITS}, HCC-TACE~\cite{HCC-TACE-Seg}, IRCAD~\cite{3D-IRCADb}. (2) Pancreas tumors: MSD07-Pancreas~\cite{msd}, PANORAMA~\cite{PANORAMA}, QUBIQ~\cite{Qubiq}. (3) Kidney tumors: KiTS21~\cite{kits}, KiTS23~\cite{kits}, KIPA~\cite{KIPA}. (4) Lung tumors: MSD06-Lung~\cite{msd}, RIDER~\cite{RIDER-LungCT-Seg}. (5) COVID-19: CV19-20~\cite{COVID}. The details of datasets are presented in \textbf{Extended Data Table~\ref{table_dataset}}. For tumor segmentation, we utilize Dice scores to measure the segmentation performance. We utilize F1-Score, sensitivity, and specificity to measure the detection performance as previous methods~\cite{biomedparse,cao2023large}.

\subsection{Clinician Evaluation}
\label{sec_result_turing}

We invited 13 board-certified radiologists to evaluate the fidelity of synthetic tumors through a Visual Turing Tests~\cite{chen2021synthetic}. These radiologists are from 4 hospitals in China, \emph{i.e.}, Li Ka Shing Faculty of Medicine of The University of Hong Kong (HKU), Shenzhen People's Hospital, Sun Yat-Sen Memorial Hospital of Sun Yat-Sen University, and The Third Affiliated Hospital of Southern Medical University. Among the group of 13 radiologists, there are 6 junior radiologists, 4 mid-level radiologists, and 3 senior radiologists. Each level of radiologists is defined by the following standards:
\begin{itemize}
    \item \textbf{Junior radiologists}: Doctors in residency programs, with 5-10 years of clinical experience.
    \item \textbf{Mid-level radiologists}: Doctors with a professional tenure of 10-20 years in hospitals.
    \item \textbf{Senior radiologists}: Doctors with advanced professional titles in hospitals, with at least 20 years of clinical experience.
\end{itemize}

The process of the Visual Turing Test is shown in \textbf{Figure~\ref{fig_overview}~(e)}. During the Visual Turing Test, 13 radiologists were presented with the same set of CT volumes containing tumors, with each volume containing only one tumor case for evaluation. 
Half of these tumors are real and the remaining half are synthesized by FreeTumor. Specifically, we provided 18 cases each of liver tumors, pancreas tumors, kidney tumors, lung tumors, and COVID-19 (a total of 90 cases) for evaluation. There are 45 real and 45 synthetic tumors among 90 tumor cases. For each type, the numbers of real and synthetic tumors are also equal (9 real and 9 synthetic in 18 tumor cases). 
These 90 cases are randomly selected from our datasets. During the Visual Turing Test, the radiologists were tasked with: \textbf{(1)} Identifying the synthetic tumors from real ones. \textbf{(2)} Discerning the distinguishing features between real and synthetic tumors. The radiologists were informed of the type of tumors they were required to identify, and the positions of tumors were also provided. The specific number of synthetic tumors for each type is unknown to the invited radiologists to prevent any bias in their assessments. On average, a radiologist would require about 2-3 hours to assess all 90 cases. 


\begin{figure*}
	\centering
	\includegraphics[width=0.9\linewidth]{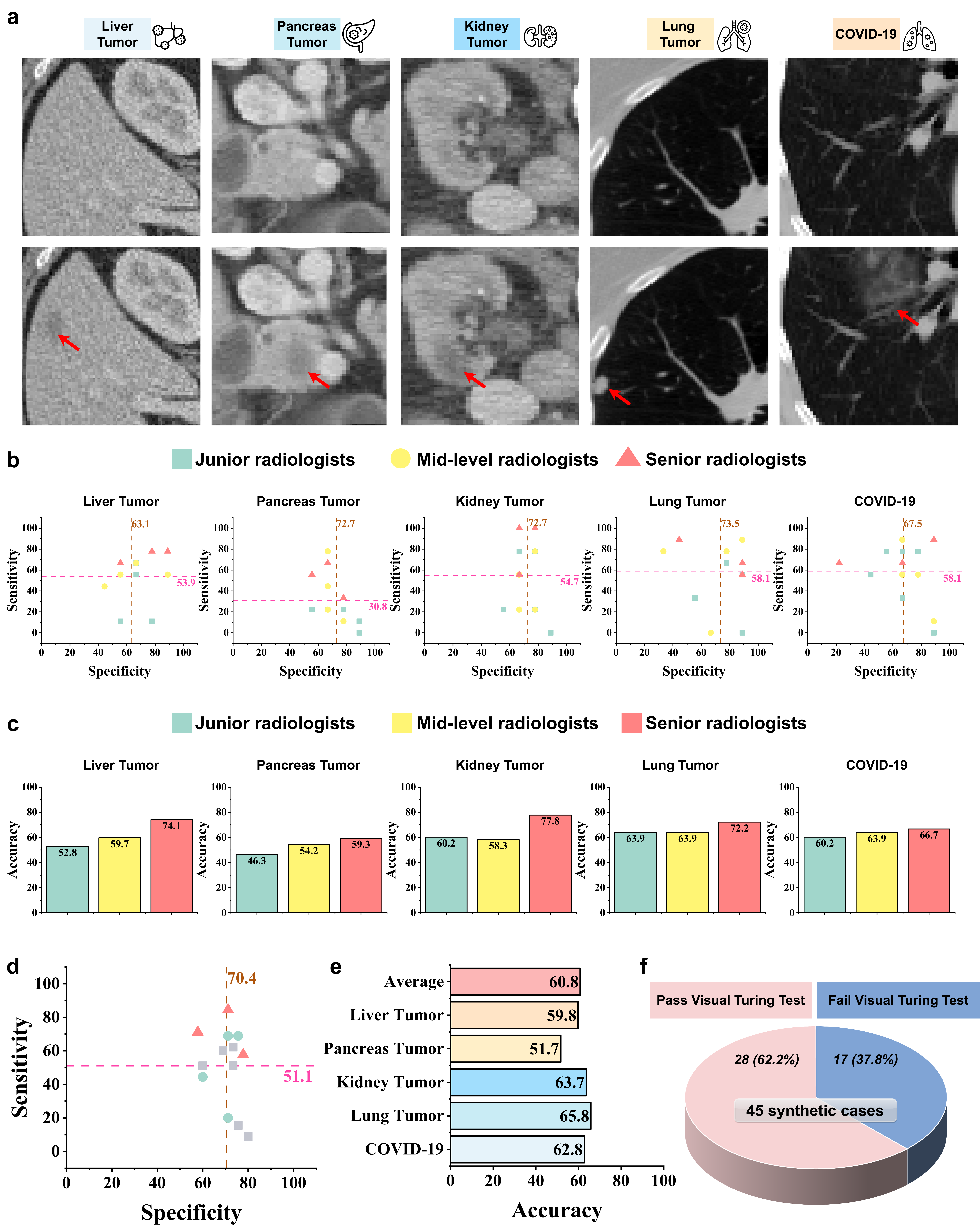}
	\caption{\textbf{Clinician evaluation of synthetic tumors}. We engage 13 board-certified radiologists in a Visual Turing Test to discern between synthetic and real tumors as shown in \textbf{Figure~\ref{fig_overview}~(e)}. \textbf{a}. Qualitative results of synthetic tumors. The upper row presents healthy organs and the lower row presents synthetic tumors on healthy organs (highlighted by \textcolor{red}{red arrows}). \textbf{b}. The sensitivity and specificity results across five types of tumors/lesions. \textbf{c}. The accuracy results across five types of tumors/lesions. \textbf{d}. The average sensitivity and specificity results across five types of tumors/lesions. \textbf{e}. The average accuracy results across five types of tumors/lesions. \textbf{f}. We divide the synthetic tumors into two groups, \emph{i.e.}, pass the Visual Turing Test and fail the Visual Turning test. Detailed results are presented in \textbf{Extended Data Tables~\ref{table_turing_test_sen}, \ref{table_turing_test_acc} and~\ref{table_average_turing}}.}
	\label{fig_turing}
\end{figure*}
\clearpage


As shown in \textbf{Figure~\ref{fig_turing}}, we report the sensitivity, specificity, and accuracy results to measure the ability of radiologists to identify our synthetic tumors. Lower values for sensitivity, specificity, and accuracy indicate that our synthetic tumors attain a higher quality level. We observe that even experienced radiologists are unable to identify our synthetic tumors with complete accuracy, which demonstrates the effectiveness of FreeTumor in synthesizing realistic tumors. Detailed results are presented in \textbf{Extended Data Tables~\ref{table_turing_test_sen}, \ref{table_turing_test_acc}, and~\ref{table_average_turing}}. Concretely:
\begin{itemize}
    \item \textbf{Sensitivity and specificity}. The sensitivity and specificity results for each type of tumor are depicted in \textbf{Figure~\ref{fig_turing}~(b)}, with the average results showcased in \textbf{Figure~\ref{fig_turing}~(d)}. Notably, the average sensitivity is recorded at a modest $\mathbf{51.1\%}$, demonstrating that FreeTumor effectively synthesizes realistic tumors.
    
    \item \textbf{Accuracy}. The accuracy results for each type of tumor are depicted in \textbf{Figure~\ref{fig_turing}~(c)}, with the average results showcased in \textbf{Figure~\ref{fig_turing}~(e)}. The accuracy results are $59.8\%$, $51.7\%$, $63.7\%$, $65.8\%$, and $62.8\%$ for liver tumors, pancreas tumors, kidney tumors, lung tumors, and COVID-19, respectively. The average accuracy of assessment is $\mathbf{60.8\%}$, suggesting that nearly $40\%$ of cases are misclassified. 
    
    \item \textbf{Junior radiologists struggle in distinguishing our synthetic tumors from real ones}. We engage radiologists of varying expertise levels to evaluate the synthetic tumors. Our observations reveal that the breadth of experience significantly influences the evaluation results. As shown in \textbf{Figure~\ref{fig_turing}~(c)}, 6 junior radiologists achieve only $\mathbf{41.5\%}$ sensitivity and $\mathbf{56.6\%}$ accuracy, indicating that our synthetic tumors exhibit realistic characteristics, capable of misleading radiologists with limited experience levels.
    
    \item \textbf{Comparisons among different types of tumors/lesions}. As shown in \textbf{Figure~\ref{fig_turing}~(e)}, among the five assessed types, pancreas tumors present the greatest challenge in identification, achieving a low sensitivity of $\mathbf{30.8\%}$.
    
    \item \textbf{Case analysis in Visual Turing Test}. Based on the results of clinician evaluation, we categorize the synthetic tumors into two groups: (1) Pass the Visual Turing Test: more than 1/2 of 13 radiologists identified the synthetic tumors as real ones. (2) Fail the Visual Turing Test: fewer than 1/2 of 13 radiologists identified the synthetic tumors as real ones. The detailed distributions of these two groups are shown in \textbf{Figure~\ref{fig_turing}~(f)}. It can be observed that there are 28 of 45 synthetic tumors ($\mathbf{62.3\%}$) pass the Visual Turing Test, indicating the high quality of our synthetic tumors.
    
\end{itemize}

\textbf{Case studies}. The case studies of synthetic tumors are presented in \textbf{Extended Data Figure~\ref{fig_Case_study1}}. Summarized from the radiologists' assessment, we highlight some characteristics of our synthetic tumors that contribute to deceiving radiologists: (1) Density: our synthetic tumors exhibit uneven and indistinct densities that are consistent with the clinical presentations of tumors. (2) Boundary: our synthetic tumors present unclear boundaries with blurred edges, resembling the characteristics of real tumors. (3) Mass Effect: our synthetic tumors also showcase the mass effect on the surrounding organs as real tumors. However, in some cases, some radiologists can still tell the distinct features of synthetic tumors, suggesting that our synthetic results can be further improved. More case studies with failure cases are presented in \textbf{Extended Data Figure~\ref{fig_Case_study2}}.

\subsection{Accurate and Scalable Segmentation across Five Types of Tumors/Lesions}
\label{sec_result_segmentation}

\textbf{Comparison methods}. We conduct extensive tumor segmentation experiments on 12 public datasets and report the corresponding Dice score results. First, we compare our FreeTumor with five widely-used tumor segmentation models~\cite{UNET,transunet,unetr,nnUNet,swin}, \emph{i.e.}, UNet~\cite{UNET}, TransUNet~\cite{transunet}, UNETR~\cite{unetr}, nnUNet~\cite{nnUNet}, and SwinUNETR~\cite{swin}. These works~\cite{UNET,transunet,unetr,nnUNet,swin} proposed to advance network architectures for improving tumor segmentation, while our FreeTumor is designed to address the challenges in tumor segmentation from the data scarcity aspect. We adopt SwinUNETR~\cite{swin} as the segmentation model, thus SwinUNETR~\cite{swin} can be seen as the baseline for comparisons. 
Second, we compare FreeTumor with two tumor synthesis methods~\cite{Syntumor,Difftumor} and three CT foundation models~\cite{MAE3D,SwinSSL,voco-v1}. In addition, we further evaluate the out-of-domain performance of FreeTumor. Out-of-domain evaluation represents transferring a model trained on a source dataset to a target dataset, \emph{i.e.}, direct inference on target datasets without fine-tuning models. 

\textbf{FreeTumor outperforms baseline tumor segmentation models}. As shown in \textbf{Figure~\ref{fig_results}}, on 12 public datasets across various types of tumors/lesions, our FreeTumor consistently outperforms five widely-used tumor segmentation models~\cite{UNET,transunet,unetr,nnUNet,swin} by a clear margin. By augmenting the training datasets by over 40 times, FreeTumor surpasses the baseline SwinUNETR~\cite{swin} by $6.9\%$, $8.6\%$, $16.1\%$, $6.0\%$, $3.1\%$, $7.2\%$, $4.0\%$, $3.7\%$, $5.8\%$, $7.1\%$, $5.1\%$, and $7.9\%$ on 12 datasets, respectively. Overall, FreeTumor brings an average $+6.7\%$ Dice score improvements (two-sided paired t-test $p$-values $<5.09\times10^{-5}$) over the baseline SwinUNETR~\cite{swin}, which is a non-trivial advancement in tumor segmentation. 
The substantial improvements demonstrate that the scarcity of tumor annotations is a critical bottleneck in tumor segmentation. Specifically, as shown in \textbf{Figure~\ref{fig_results}~(c)}, for the IRCAD~\cite{3D-IRCADb} dataset that contains only 22 labeled CT volumes, FreeTumor demonstrates $\mathbf{+16.1\%}$ Dice score improvements by augmenting training datasets. These findings robustly validate the rationale of our motivation to mitigate data scarcity. Detailed results are presented in \textbf{Extended Data Table~\ref{table_results}}.

\textbf{FreeTumor outperforms previous tumor synthesis methods}. We further compare FreeTumor with two tumor synthesis methods: SynTumor~\cite{Syntumor} and DiffTumor~\cite{Difftumor}. Note that both of these two tumor synthesis methods~\cite{Syntumor, Difftumor} cannot leverage unlabeled data for synthesis training: (1) SynTumor~\cite{Syntumor} utilizes handcrafted image processing techniques for tumor synthesis. (2) DiffTumor~\cite{Difftumor} employs conditioned diffusion models for tumor synthesis, thus it can only leverage labeled data for tumor synthesis training (360 labeled volumes are used in this work). 
In addition, SynTumor~\cite{Syntumor} is only applicable to liver tumors, and DiffTumor~\cite{Difftumor} is not applicable to lung tumors and COVID-19. For fair comparisons, SynTumor~\cite{Syntumor} and DiffTumor~\cite{Difftumor} adopt the same segmentation model~\cite{swin} as FreeTumor. 

\begin{figure*}
	\centering
	\includegraphics[width=0.95\linewidth]{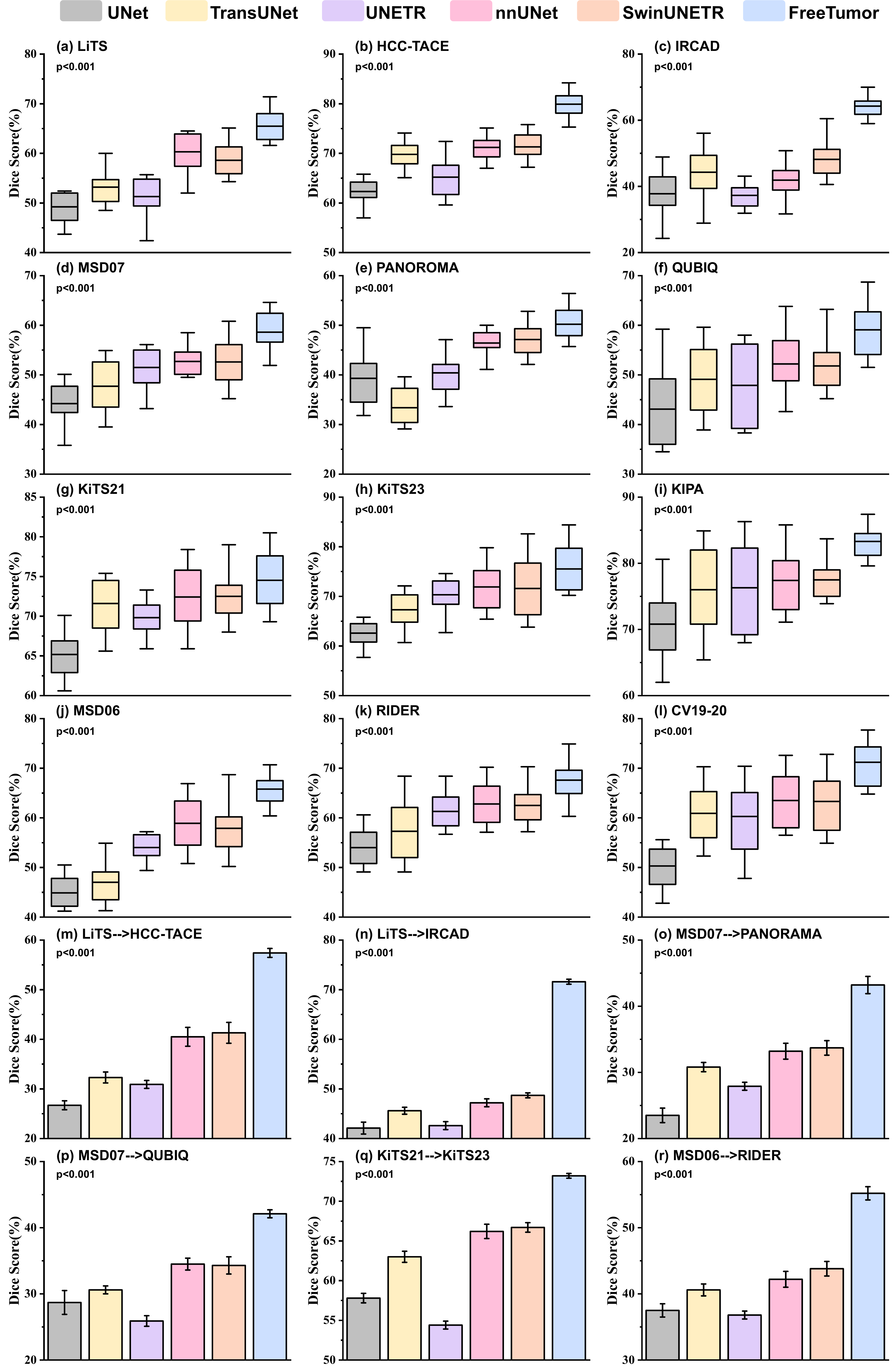}
\end{figure*}
\clearpage
\begin{figure*}
	\caption{\textbf{Comparison with baseline tumor segmentation models}. \textbf{a-l.} The 5-fold cross-validation results of 12 public datasets. Specifically, FreeTumor adopts SwinUNETR~\cite{swin} as the segmentation model for segmentation. Overall, FreeTumor brings an average $+6.7\%$ Dice score improvements (two-sided paired t-test $p$-values $<5.09\times10^{-5}$) over the baseline SwinUNETR~\cite{swin}. \textbf{m-r.} Out-of-domain evaluation. The standard deviations are obtained from five times of experiments. Specifically, we train the model on a source dataset and conduct direct inference on a target dataset without fine-tuning. For example, in (m), ``LiTS to HCC-TACE'' represents training a model on the LiTS~\cite{LITS} dataset and conducting inference on the HCC-TACE~\cite{HCC-TACE-Seg} dataset without fine-tuning. Compared with the baseline SwinUNETR~\cite{swin}, FreeTumor brings average $+12.3\%$ Dice score improvements (two-sided paired t-test $p$-values $<4.42\times10^{-3}$) in 6 out-of-domain experiments. Detailed results are presented in \textbf{Extended Data Tables~\ref{table_results} and~\ref{table_out_of_domain}}.}
	\label{fig_results}
\end{figure*}

As shown in \textbf{Figure~\ref{fig_results_syn_ssl}}, FreeTumor significantly outperforms previous tumor synthesis methods SynTumor~\cite{Syntumor} and DiffTumor~\cite{Difftumor} by a clear margin, underscoring the importance of leveraging large-scale data for synthesis training.
We further evaluate the effectiveness of SynTumor~\cite{Syntumor} and DiffTumor~\cite{Difftumor} in utilizing our large-scale datasets for segmentation training. However, we observe that without large-scale synthesis training, these synthesis methods~\cite{Syntumor,Difftumor} fail to generalize well on large-scale unseen datasets with different image characteristics. 
For example, when employing SynTumor~\cite{Syntumor} to segmentation training on our large-scale datasets, the average Dice score on LiTS~\cite{LITS} is dropped from $60.2\%$ to $52.8\%$. 
Detailed results are presented in \textbf{Extended Data Table~\ref{table_abla_scale_other_synthesis} and Figure~\ref{fig_vsDiff}}. In contrast, our FreeTumor is capable of leveraging large-scale data in both synthesis and segmentation training, facilitating robust generalization across datasets from various sources. Detailed results are presented in \textbf{Extended Data Table~\ref{table_compare_synthesis}}.

\textbf{FreeTumor outperforms various CT foundation models}. We further compare FreeTumor with three CT foundation models: MAE3D~\cite{MAE3D}, SwinSSL~\cite{SwinSSL}, and VoCo~\cite{voco-v1}. These foundation models are based on Self-Supervised Learning (SSL)~\cite{MAE,MOCO,simclr}: MAE3D~\cite{MAE3D} and SwinSSL~\cite{SwinSSL} are based on mask image modeling~\cite{MAE}, while VoCo~\cite{voco-v1} is based on contrastive learning. Although these foundation models~\cite{MAE3D,SwinSSL,voco-v1} can leverage unlabeled data in self-supervised pre-training, they still fail to utilize unlabeled data during segmentation training and remain constrained by the limited scale of annotated datasets. 

As shown in \textbf{Figure~\ref{fig_results_syn_ssl}}, we observe that our FreeTumor clearly outperforms three foundation models~\cite{MAE3D,SwinSSL,voco-v1}.  
The fundamental bottleneck of the foundation models~\cite{MAE3D,SwinSSL,voco-v1} is that they fail to leverage large-scale data during segmentation training.
For example, for the liver tumor dataset IRCAD~\cite{3D-IRCADb}, these foundation models~\cite{MAE3D,SwinSSL,voco-v1} are limited to utilizing merely \textbf{22} CT volumes for fine-tuning, whereas our FreeTumor model can harness a significantly larger dataset of \textbf{19,571} CT volumes for segmentation training. The utilization of large-scale data in segmentation training enables the superiority of FreeTumor. Detailed results are presented in \textbf{Extended Data Table~\ref{table_compare_ssl}}.

\begin{figure*}[!h]
	\centering
	\includegraphics[width=0.95\linewidth]{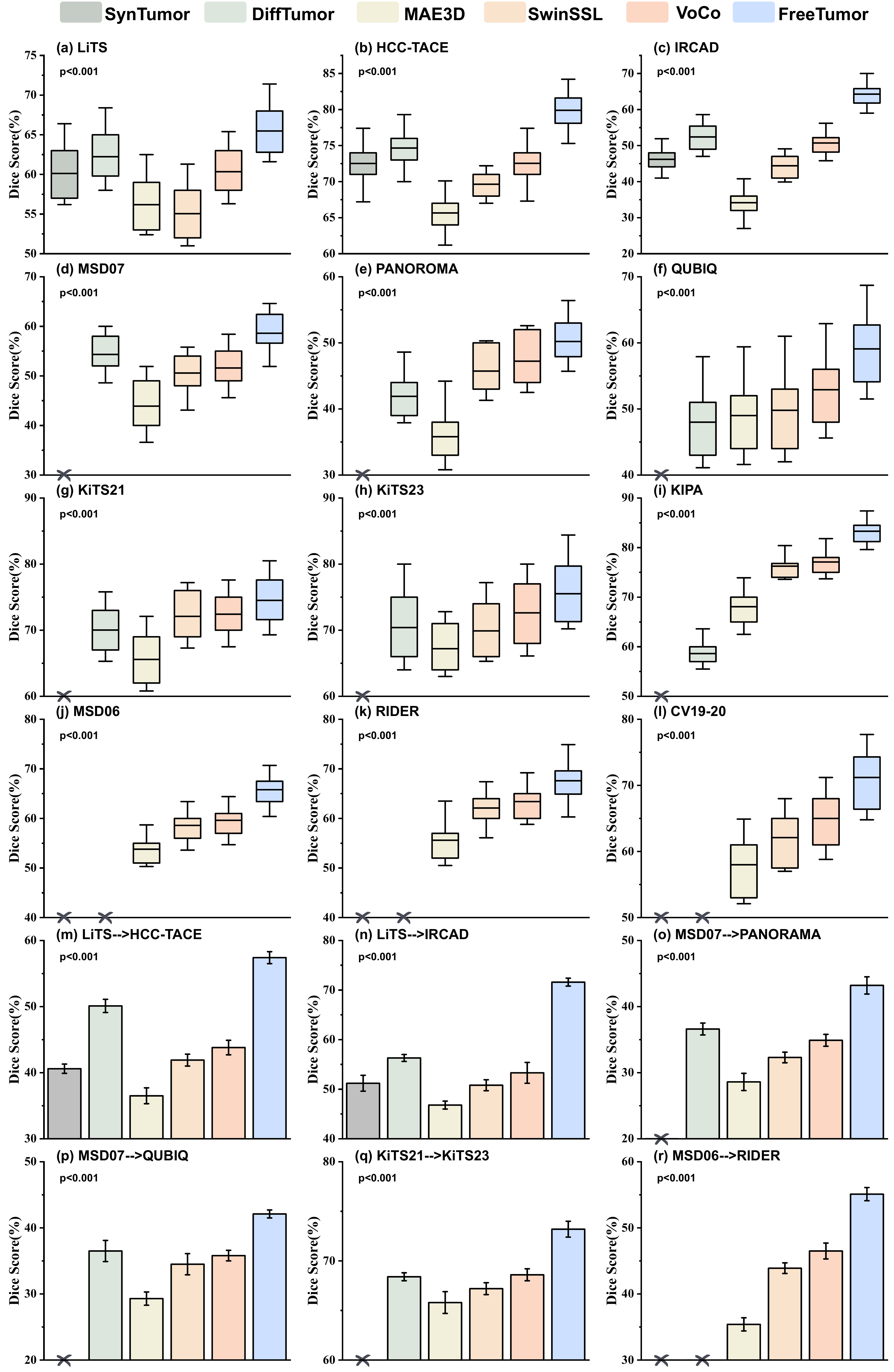}
\end{figure*}
\clearpage
\begin{figure*}
	\caption{\textbf{Comparison with tumor synthesis methods and CT foundation models}. \textbf{a-l.} The 5-fold cross-validation results of 12 public datasets. SynTumor~\cite{Syntumor} and DiffTumor~\cite{Difftumor} are two tumor synthesis methods using the same segmentation model~\cite{swin} as FreeTumor, while SynTumor~\cite{Syntumor} is only applicable to liver tumors, and DiffTumor~\cite{Difftumor} is not applicable to lung tumors and COVID-19. We use a ``cross mark'' (\XSolidBrush) to signify that this method is not applicable to this dataset. For example, the ``cross mark'' in (d) means SynTumor~\cite{Syntumor} is not applicable to the pancreas tumor dataset MSD07~\cite{msd}. In addition, MAE3D~\cite{MAE}, SwinSSL~\cite{SwinSSL}, and VoCo~\cite{voco-v1} are three CT foundation models based on self-supervised learning. The same segmentation model~\cite{swin} is adopted for fair comparisons. Overall, on 12 public datasets, FreeTumor surpasses the best-competing method by average $5.1\%$ Dice scores (two-sided paired t-test $p$-values $< 3.78 \times 10^{-5}$). \textbf{m-r.} Out-of-domain evaluation. The standard deviations are obtained from five times of experiments. Overall, in 6 out-of-domain experiments, FreeTumor surpasses the best-competing method by average $7.9\%$ Dice scores (two-sided paired t-test $p$-values $<3.73\times10^{-3}$) in out-of-domain evaluation. Detailed results are presented in \textbf{Extended Data Tables~\ref{table_compare_synthesis}, \ref{table_compare_ssl}, and~\ref{table_out_of_domain}}.}
	\label{fig_results_syn_ssl}
\end{figure*}

\textbf{FreeTumor excels in out-of-domain evaluation}. Extensive out-of-domain comparisons with five tumor segmentation models~\cite{UNET, transunet, unetr, nnUNet, swin}, two tumor synthesis methods~\cite{Syntumor,Difftumor}, and three foundation models~\cite{MAE3D,SwinSSL,voco-v1} are presented in \textbf{Figure~\ref{fig_results}~(m-r) and Figure~\ref{fig_results_syn_ssl}~(m-r)}, respectively. Leveraging large-scale data from diverse sources, FreeTumor demonstrates superior generalizability compared with previous methods. Notably, when transferring models from LiTS~\cite{LITS} to IRCAD~\cite{3D-IRCADb}, FreeTumor achieves a substantial improvement of $\mathbf{22.9\%}$ Dice score compared with the baseline SwinUNETR~\cite{swin} and also surpasses both tumor synthesis methods~\cite{Syntumor,Difftumor}, and foundation models~\cite{MAE3D,SwinSSL,voco-v1} by a clear margin. Detailed results are presented in \textbf{Extended Data Table~\ref{table_compare_synthesis}}. 

\textbf{FreeTumor yields significant improvements across five types of tumors/lesions}. As shown in \textbf{Figure~\ref{fig_segmentation}~(a)}, compared with the baseline SwinUNETR~\cite{swin}, FreeTumor yields average improvements of $10.6\%$, $5.5\%$, $3.8\%$, $6.1\%$, and $7.9\%$ for liver tumors, pancreas tumors, kidney tumors, lung tumors, and COVID-19 in Dice scores, respectively. Given the marginal disparities observed within previous methods~\cite{UNET, transunet, unetr, nnUNet, Syntumor,Difftumor, swin,MAE3D,SwinSSL,voco-v1}, these improvements underscore a non-trivial advancement in tumor segmentation.
We provide qualitative visualization results of tumor segmentation in \textbf{Figure~\ref{fig_segmentation}~(c)}. Notably, FreeTumor demonstrates outstanding segmentation performance, offering precise sizes, shapes, and positions that are crucial for accurate tumor diagnosis. More qualitative results are presented in \textbf{Extended Data Figure~\ref{fig_comparison}}.


\begin{figure*}
	\centering
	\includegraphics[width=1\linewidth]{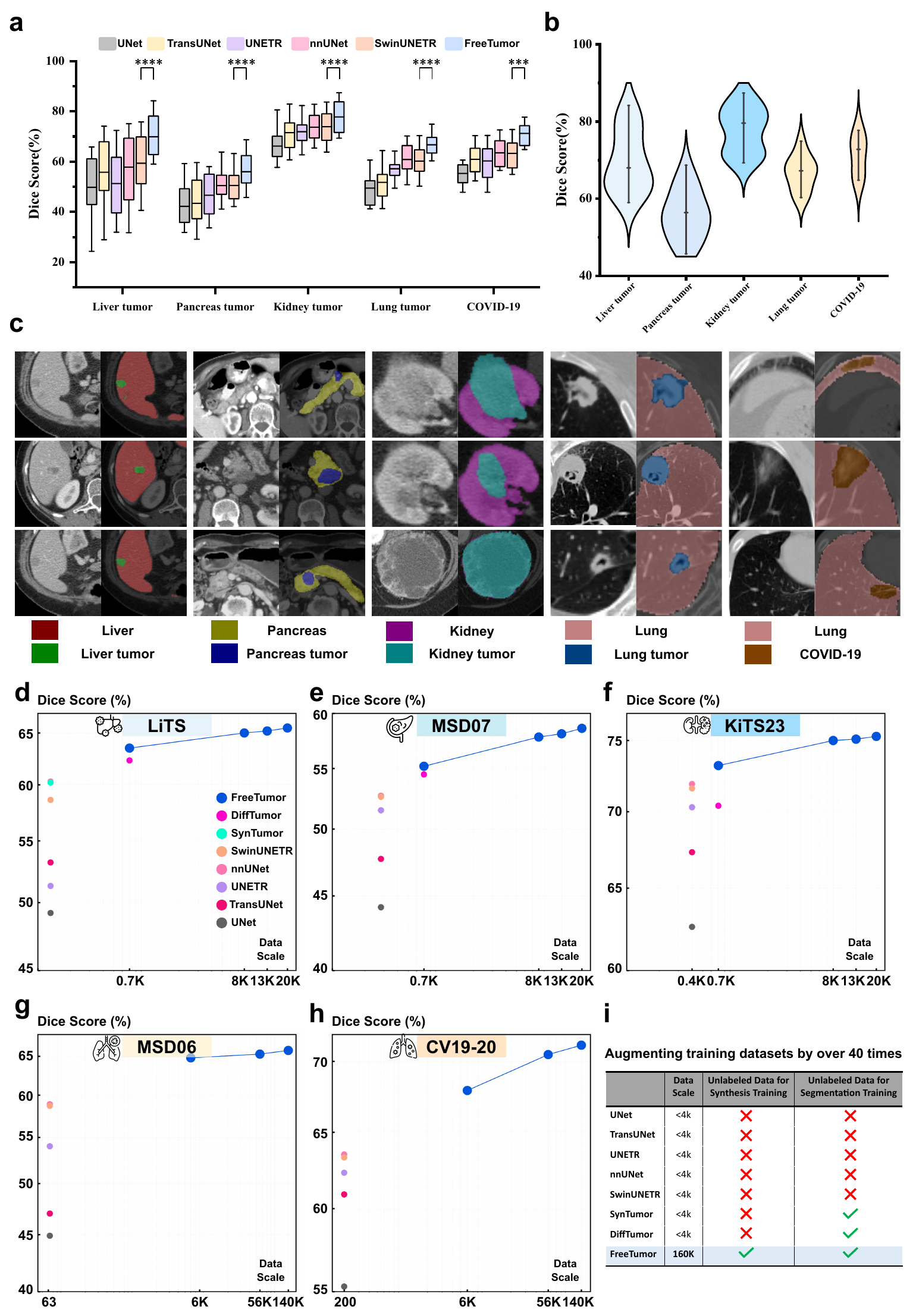}
\end{figure*}
\clearpage
\begin{figure*}
	\caption{\textbf{Comprehensive analysis of tumor segmentation performance and data scaling effects}. \textbf{a}. The overall Dice score comparisons with baseline tumor segmentation models~\cite{UNET,transunet,unetr,nnUNet,swin}. Significance levels at which FreeTumor outperforms the baseline SwinUNETR~\cite{swin}, with two-sided paired t-test are ***$p$-values $< 1 \times 10^{-3}$ and ****$p$-values $< 1 \times 10^{-4}$. Exact $p$-values for the comparison between FreeTumor and SwinUNETR~\cite{swin} are: $p$-values $< 6.05 \times 10^{-7}$ for liver tumors, $p$-values $< 4.02 \times 10^{-7}$ for pancreas tumors, $p$-values $< 1.05 \times 10^{-5}$ for kidney tumors, $p$-values $< 7.37 \times 10^{-5}$ for lung tumors, and $p$-values $< 9.07 \times 10^{-4}$ for COVID-19. \textbf{b}. The average Dice scores of FreeTumor across five types of tumors/lesions. \textbf{c}. Qualitative segmentation results of FreeTumor. The organ segmentation results are presented for better visualization. \textbf{d-h}. The effectiveness of scaling up training datasets. We evaluate the correlation between the data scale of segmentation training datasets and segmentation performances. Specifically, the foundation models~\cite{MAE3D,SwinSSL,voco-v1} are unable to utilize unlabeled data in segmentation training, thus their data scale of segmentation training datasets are the same as the baseline models~\cite{UNET,transunet,unetr,nnUNet,swin}.  \textbf{i}. Comparisons between FreeTumor and previous methods~\cite{UNET,transunet,unetr,nnUNet,swin,Syntumor,Difftumor} in data utilization. We assess these methods across three dimensions: the scale of training datasets (number of CT volumes), the utilization of unlabeled data in synthesis training, and the utilization of unlabeled data in segmentation training.}
	\label{fig_segmentation}
\end{figure*}

\textbf{Large-scale data enables more accurate tumor segmentation}. The key strength of FreeTumor lies in its capacity to harness large-scale unlabeled data for tumor synthesis and segmentation. As shown in \textbf{Figure~\ref{fig_segmentation}~(d-h)}, we showcase the effectiveness of scaling up segmentation training datasets across five segmentation datasets~\cite{LITS,msd,kits,COVID}, representing five types of tumors/lesions.
We present the comparisons with five baseline models~\cite{UNET,transunet,unetr,nnUNet,swin} and two tumor synthesis methods~\cite{Syntumor,Difftumor}. 
The foundation models~\cite{MAE3D, SwinSSL, voco-v1} leveraged segmentation training datasets that are of equivalent scale to the baseline models~\cite{UNET, transunet, unetr, nnUNet, swin}.


We have noted a significant correlation between segmentation performance and the scale of segmentation training datasets. As shown in \textbf{Figure~\ref{fig_segmentation}~(i)}, we further present a comparative analysis of data utilization. Notably, a key distinction lies in the utilization of unlabeled data. Previous methods~\cite{UNET, transunet, unetr, nnUNet, swin, Syntumor, Difftumor} are limited to less than 4,000 CT volumes for training. 
Although two previous methods SynTumor~\cite{Syntumor} and DiffTumor~\cite{Difftumor} also explore tumor synthesis, they are unable to leverage large-scale unlabeled data for synthesis training. 
Without synthesis training on large-scale data, these two synthesis methods~\cite{Syntumor,Difftumor} fall short in effectively leveraging large-scale data for segmentation training (\textbf{Extended Data Table~\ref{table_abla_scale_other_synthesis}}).
In summary, previous methods~\cite{UNET, transunet, unetr, nnUNet, swin, Syntumor,Difftumor} are constrained by their reliance on limited labeled data, thus curbing their potential for achieving superior performances. In contrast, by integrating large-scale data for tumor synthesis and segmentation training, our FreeTumor surpasses previous methods~\cite{UNET, transunet, unetr, nnUNet, swin, Syntumor, Difftumor} by a clear margin. These findings unequivocally demonstrate the rationale and effectiveness of FreeTumor.

\subsection{Accurate Detection across Five Types of Tumors/Lesions}
\label{sec_result_early}


 
\begin{figure*}
	\centering
	\includegraphics[width=0.85\linewidth]{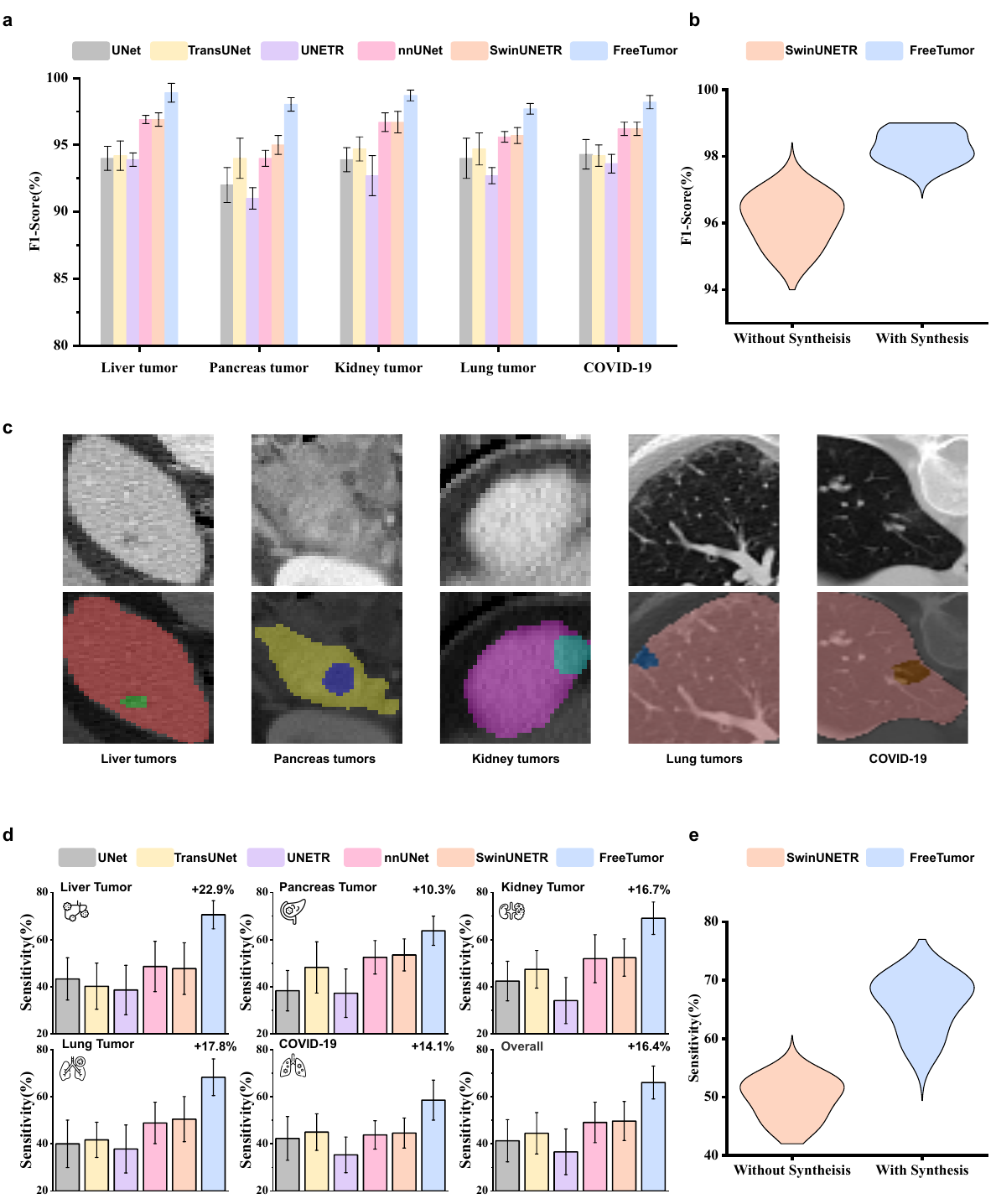}
	\caption{\textbf{Evaluation of tumor detection}. \textbf{a.} The overall detection performances of all stages of tumors/lesions. We report the results in terms of F1-Score. \textbf{b.} The average F1-Score results of detecting five types of tumors/lesions. ``Without synthesis'' represents the F1-Score results of the baseline SwinUNETR~\cite{swin} model for comparison. With tumor synthesis, FreeTumor yields average $+2.3\%$ F1-Score improvements (two-sided paired t-test $p$-values $<4.31\times10^{-4}$). \textbf{c.} Qualitative visualization results of detecting small tumors/lesions. \textbf{d.} The sensitivity results of detecting small tumors/lesions (diameter $<$ 2cm). \textbf{e.} The average sensitivity results of detecting five types of small tumors/lesions. ``Without synthesis'' represents the sensitivity results of the baseline SwinUNETR~\cite{swin} model. With tumor synthesis, FreeTumor yields average $+16.4\%$ sensitivity improvements (two-sided paired t-test $p$-values $<1.45\times10^{-3}$) in detecting small tumors/lesions. Detailed results are presented in \textbf{Extended Data Table~\ref{table_early}}.}
	\label{fig_early}
\end{figure*}

Tumor detection, especially the detection of early-stage tumors, is vital for the timely treatment of patients. Accurate early tumor detection can result in a greater probability of survival with less morbidity as well as less expensive treatment~\cite{choi2014ct,cao2023large,fitzgerald2022future,singhi2019early,pereira2020early}. However, early-stage tumors are typically small in size, making them challenging to detect. Our proposed FreeTumor can synthesize tumors with flexible sizes.
Thus, the synthesis of small tumors can serve as an effective data augmentation solution to improve the robustness of early tumor detection. In this study, we employ FreeTumor to synthesize a large number of small tumors for training, thereby boosting the sensitivity of early tumor detection and facilitating the timely treatment for patients.

\textbf{Evaluation of tumor detection across all stages of tumors}. We first evaluate the detection performance across all tumor stages, with the F1-Score ($\%$) results illustrated in \textbf{Figure~\ref{fig_early}~(a)}. 
It can be seen that FreeTumor consistently surpasses the baseline methods~\cite{UNET,transunet,unetr,nnUNet,swin} without tumor synthesis. Notably, the F1-Scores of FreeTumor in detecting the five types of tumors/lesions all surpass $97\%$, highlighting the potential of FreeTumor in clinical practice.


\textbf{Effectiveness of detecting small tumors}. To evaluate the performances of early tumor detection, we further present the results of detecting small tumors (diameter $<$ 2cm)~\cite{miller1981reporting,bassi2025radgpt}. We highlight the sensitivity improvements of FreeTumor in \textbf{Figure~\ref{fig_early}~(d)}. It can be seen that limited by the data scarcity, the baseline methods~\cite{UNET,transunet,unetr,nnUNet,swin} are not sensitive in detecting small tumors/lesions. Equipped with FreeTumor, the detection of small liver tumors, pancreas tumors, kidney tumors, lung tumors, and COVID-19 are improved by $22.9\%$, $10.3\%$, $16.7\%$, $17.8\%$, and $14.1\%$, respectively. Notably, the overall sensitivity is improved from $49.7\%$ to $66.1\%$ ($\mathbf{+16.4\%}$), marking a substantial advancement towards accurate early tumor detection. These findings indicate promising prospects of FreeTumor in aiding the timely treatment of patients. Detailed sensitivity and specificity results are presented in \textbf{Extended Data Figure~\ref{fig_early_sen_spe}}.



\section{Discussion}\label{sec3}

\textbf{FreeTumor is the first GAI framework tailored for large-scale tumor synthesis and segmentation training}. Our FreeTumor is designed to address the scarcity of annotated tumor datasets, aiming to unleash the power of large-scale unlabeled data for training. Specifically, FreeTumor effectively leverages a combination of limited labeled data and large-scale unlabeled data for tumor synthesis training. By large-scale tumor synthesis training, FreeTumor is capable of synthesizing a large number of tumors varying in sizes, positions, and backgrounds, thus boosting the robustness of tumor recognition models. Rigorous clinician evaluation conducted by 13 board-certified radiologists demonstrates the high quality of our synthetic tumors. 
To evaluate the effectiveness of FreeTumor, we create the largest training dataset for tumor synthesis and recognition, encompassing 161,310 publicly available CT volumes from diverse sources (with only $2.3\%$ of them containing annotated tumors). Extensive experiments on 12 public datasets demonstrate the superiority of FreeTumor over state-of-the-art AI methods. These findings showcase the promising prospects of FreeTumor in tumor recognition. 

AI-driven tumor recognition has received increasing attention in recent years, yet the progress is heavily hampered by the scarcity of annotated datasets. Early attempts~\cite{UNET,transunet,unetr,nnUNet,swin} mainly focus on advancing network architectures to improve tumor recognition. Although encouraging results have been demonstrated, the scarcity of annotated datasets still heavily hampered further development. 
To this end, numerous medical foundation models~\cite{MAE3D,SwinSSL,voco-v1,zhuang2024mim,ni2024mg} have been introduced to tackle the challenges of data scarcity. Although these foundation models can leverage unlabeled data in self-supervised pre-training~\cite{oquab2024dinov2,caron2021emerging, wu2023sparsely,wu2024modeling,wu2022deep,wu2022DCA,liu2023multi, MOCO,simclr,MAE}, they still fail to utilize unlabeled data during segmentation training and remain constrained by the limited scale of annotated datasets. 

Thus, tumor synthesis emerges as a promising solution to mitigate the scarcity of annotated tumor datasets, which can synthesize a large number of tumors on images for augmenting training datasets. Early attempts~\cite{Syntumor,lyu2022pseudo,yao2021label,wang2022anomaly,wyatt2022anoddpm,Difftumor} investigated image processing and generative models for tumor synthesis. However, these methods fail to integrate large-scale data into synthesis training, thus hindering the improvements of downstream tumor recognition. In addition, these methods largely ignore the importance of quality control in synthesizing tumors, while low-quality synthetic tumors will pose a negative impact on downstream training.

To this end, we introduce FreeTumor to address the aforementioned challenges. First, FreeTumor adopts a novel adversarial-based synthesis training framework to leverage both labeled and unlabeled data, facilitating the integration of large-scale unlabeled data in synthesis training. Second, FreeTumor further employs an adversarial-based discriminator to discard low-quality synthetic tumors, enabling automatic quality control of large-scale synthetic tumors in the subsequent segmentation training. In this way, FreeTumor facilitates the utilization of large-scale data in both synthesis and segmentation training, demonstrating superior performances compared with previous methods.

Although FreeTumor has demonstrated promising results in tumor recognition, there are still numerous areas for growth and improvement. 
FreeTumor has showcased promising results in synthesizing various types of tumors/lesions on CT volumes. Moving forward, we aim to extend the application of FreeTumor to encompass other tumor types and other medical imaging modalities (\emph{e.g.}, Magnetic Resonance Imaging and pathology images). 
In addition, while FreeTumor has achieved satisfactory performance on various public datasets, further exploration of its application in clinical practice is necessary to substantiate the effectiveness of our method.

\section{Methods}
\label{sec_method}

In this section, we first introduce the preliminary of our method in Section~\ref{sec_method_Preliminary}. The details of our tumor synthesis pipeline are illustrated in Section~\ref{sec_method_synthesis}. Then, in Section~\ref{sec_method_filter}, we further describe our quality control strategy to discard low-quality synthetic tumors. Following this, in Section~\ref{sec_method_unleash}, we discuss the process of integrating large-scale unlabeled data in segmentation training. Finally, in Section~\ref{sec_method_implementation}, we delve into the details of our implementation, including the details of dataset collection, pre-processing, training implementations, and evaluation metrics.

In this study, we focus on the tumor recognition tasks, thus we use the term ``unlabeled'' to represent ``without tumor labels''. Specifically, during tumor synthesis, we require organ labels to simulate the tumor positions on healthy organs. Among the datasets collected in this study, only a few of them contain organ labels. For the datasets that are without organ labels, we first utilize an organ segmentation model to generate pseudo organ labels. The details of pre-processing datasets are described in Section~\ref{sec_method_implementation}.

\subsection{Preliminary of FreeTumor}
\label{sec_method_Preliminary}

Confronted with the challenge of conditioned diffusion models lacking the ability to leverage unlabeled data in synthesis training~\cite{Difftumor}, we explore the adversarial training method to unleash the power of large-scale unlabeled data. Specifically, our synthesis training pipeline is motivated by the GAN-based semantic image synthesis methods~\cite{SPADE,dong2017semantic,oasis,freemask,tan2021diverse,liu2019learning,tan2021efficient}. Semantic image synthesis aims to generate images with specific classes.
Typically, GAN-based semantic image synthesis methods first train a classification model as the discriminator in the generative model. During synthesis training, this discriminator is utilized to classify the images generated by the generator, where higher classification accuracy indicates higher quality of synthetic images. In this way, the generator can be trained by minimizing the classification loss.

In this paper, we propose to shift this paradigm to the field of tumor synthesis. Specifically, instead of using classification models, we propose to train a tumor segmentation model as the discriminator to distinguish synthetic tumors. Furthermore, unlike previous semantic image synthesis methods focused solely on image generation, our synthetic tumors are utilized to augment segmentation training datasets. Thus, to alleviate the negative impact of low-quality synthetic tumors, we further leverage the discriminator to enable automatic quality control of synthetic tumors. The framework of FreeTumor is shown in \textbf{Extended Data Figure~\ref{fig_frame}}.

\subsection{Large-Scale Generative Tumor Synthesis Training}
\label{sec_method_synthesis}

First, we train a tumor segmentation model to discriminate between real and synthetic tumors. In Stage 1, we train a baseline segmentation model with only labeled tumor datasets, which will be employed as the discriminator of the following tumor synthesis model to discriminate the synthetic tumors. 

Second, we employ the adversarial training strategy to train a tumor synthesis model. The first step is to simulate the tumor positions on the healthy organs, which aims to select a proper location for the synthetic tumors. Specifically, we first generate organ labels for these datasets (as described in \textbf{Section~\ref{sec_method_implementation}}).
With organ labels, it is easy to select a location to synthesize tumors, \emph{e.g.}, liver tumors on livers, pancreas tumors on pancreases. Here, we denote the tumor mask as $M$ that represents the positions of synthetic tumors, where $M=1$ are the positions of synthetic tumors and $M=0$ remain as the original values. \textbf{The tumor mask $M$ is generated with flexible sizes and positions, enabling us to synthesize diverse tumors for boosting the robustness of tumor segmentation models}.

The generator $G$ used in this study is a typical encoder-decoder based U-Net~\cite{UNET}, which is widely-used in state-of-the-art generative models~\cite{freestyle,oasis,Difftumor,ldm}. In FreeTumor, we aim to use the generator $G$ to transform the voxel values from organ to tumor. Specifically, we use $x$ to denote the original voxel values, $\hat{x}$ denotes the synthetic voxel values, and the transform process is as follows:
\begin{equation}\label{eq_synthetic}
    \hat{x} = (1-M){\otimes}x + M{\otimes}[x - tanh(G(x)){\otimes}g(x)],
\end{equation}
where $x$ is first normalized to $0{\sim}1$ and $g(x)$ is the Gaussian filter to blur the textures, enabling us to simulate diverse tumor textures. $tanh$ is the activation function to normalize $G(x)$. With the tumor mask $M$, only the synthetic positions are transformed and other positions are reserved as the original values. According to Equation~\ref{eq_synthetic}, FreeTumor synthesizes tumors by estimating the distance ($tanh(G(x))$) between organs and tumors. This approach transforms tumor synthesis into a trainable process, enhancing its adaptability and effectiveness. 

In FreeTumor, we propose to employ a tumor segmentation model as the discriminator for adversarial training. 
During synthesis training, we feed the volumes with synthetic tumors to the segmentation model $S$. 
We aim to use the segmentation results of these synthetic tumors to optimize the generator $G$ by adversarial training. Concretely, it is intuitive that if a case of synthetic tumor appears realistic in comparison to the real tumors, it has a higher probability of being segmented by the segmentation model $S$. Similar observations are also witnessed in previous semantic image synthesis methods~\cite{SPADE,Syntumor,freemask,freestyle}.
Motivated by this, we calculate the segmentation loss $L_{seg}$ for adversarial training as follows:
\begin{equation}\label{loss_seg}
    L_{seg} = \frac{1}{\parallel M\parallel}\sum_{M=1}{\parallel 1 - S(\hat{x})\parallel},
\end{equation}
where $S(\hat{x})$ is the tumor prediction logits generated by the baseline segmentation model $S$, and we employ the simplest Euclidean distance to optimize the generator $G$. 

In addition, following the traditional GAN~\cite{zhu2017unpaired,isola2017image,SPADE,oasis}, besides the segmentation model, we also adopt another classifier discriminator $C$ to discriminate real or fake tumors using a typical classification loss $L_{cls}$. The classifier $C$ works similarly to the previous adversarial training methods: (1) In the discriminating process, $C$ is optimized to distinguish real and synthetic tumors. (2) In the generating process, $C$ is frozen and tries to classify the synthetic tumors as the real tumors, thus optimizing the generator $G$. Thus, the total adversarial training loss $L_{adv}$ is as follow:
\begin{equation}\label{loss_adv}
    L_{adv} = \mathop{max}_{G^{\sim}}\mathop{min}_{D^{\sim}}\lambda_{cls}L_{cls} + {L_{seg}},
\end{equation}
where $G^{\sim}$ and $D^{\sim}$ represent the generating and discriminating processes, respectively. $\lambda_{cls}$ is the weight of $L_{cls}$ and is set to $0.1$ in experiments empirically. Ablation studies of loss functions are presented in \textbf{Extended Data Table~\ref{table_abla_adversarial}}.

\subsection{Quality Control of Synthetic Tumors for Large-Scale Segmentation Training}
\label{sec_method_filter}
It is worth noting that synthetic tumors are not always flawless or perfect. We observe that the low-quality synthetic tumors will deteriorate the tumor segmentation training. Previous tumor synthesis methods~\cite{Syntumor,lyu2022pseudo,wang2022anomaly,wyatt2022anoddpm,Difftumor} largely ignored to alleviate their negative impacts. Thus, based on our discriminator, we develop an effective quality control strategy to automatically discard low-quality synthetic tumors.

\textbf{Segmentation-based discriminator for quality control}. Our quality control strategy relies on the segmentation-based discriminator $S$, which is a key factor in our decision to utilize adversarial training rather than diffusion models for tumor synthesis. We propose to adaptively discard low-quality synthetic tumors by calculating the proportions of satisfactory synthesized tumor regions. The satisfactory synthesized tumors represent the synthetic tumors that do match the corresponding tumor masks $M$ well. Intuitively, we can use the baseline segmentation model $S$ to calculate the correspondence: the proportions of synthetic tumors that are segmented as tumors. Thus, we calculate the proportion $P$ as follows:
\begin{equation}\label{eqn_proportion}
    P = \frac{\sum_{i=1}^{N}[\mathbbm{1}(S({\hat{x}})){\times}\mathbbm{1}(M=1))]}{\sum_{i=1}^{N}[\mathbbm{1}(M=1))]},
\end{equation}
where $N$ denotes the total number of voxels, $\mathbbm{1}(S({\hat{x}})$ denotes the number of voxels that are segmented as tumors, $\mathbbm{1}(M=1))$ denotes the number of voxels that tumor mask is $1$ (the positions of synthetic tumors). It is intuitive that \emph{if the proportion} $P$ \emph{is higher, the quality of this case of synthetic tumor tends to be higher}. In this way, the discriminator can serve as an automatic tool for quality control.

We set a threshold $T$ to split the high- and low-quality synthetic tumors. We use the term ``Quality Test" to represent whether the synthetic case passes the discriminator, the quality control strategy $Q$ is defined as:
\begin{equation}\label{eqn_turing}
Q(x|P, T, G, S)=\left\{
     \begin{array}{lr}
     \hat{x}, ~~P{\geq}T,  &\text{This synthetic tumor pass the Quality Test}\\
     x, ~~P{<}T,  &\text{This synthetic tumor fail the Quality Test}
     \end{array}
\right.
\end{equation}
With $Q$, we can effectively achieve quality control of the synthetic tumors online. Ablation studies are presented in \textbf{Extended Data Table~\ref{table_abla_filtering}}. Despite its simplicity, we effectively alleviate the negative impact of unsatisfactory synthetic tumors in segmentation training, which is a significant improvement upon the previous tumor synthesis methods~\cite{Syntumor,lyu2022pseudo,wang2022anomaly,wyatt2022anoddpm,Difftumor}. 

\subsection{Unleashing the Power of Large-scale Unlabeled Data}
\label{sec_method_unleash}

Distinguished from previous works~\cite{Syntumor,Difftumor,UNET,transunet,unetr,swin,nnUNet,voco-v1,MAE3D,SwinSSL} that used a limited scale of dataset for tumor segmentation training, we emphasize the importance of large-scale unlabeled data in the development of tumor segmentation. With the rapid development of medical imaging, we can easily collect adequate unlabeled CT data for training our FreeTumor. The challenge is that these datasets lack annotated tumor cases. To this end, we develop FreeTumor to leverage these unlabeled data.
Specifically, as described in \textbf{Sections~\ref{sec_method_synthesis} and~\ref{sec_method_filter}}, given the unlabeled datasets $D_{u}$, we conduct tumor synthesis for $D_{u}^{'}$ as follow:
\begin{equation}\label{eqn_syn_unlabeled}
    D_{u}^{'} = \left\{(x, F[G(x)], S) | x{\in}{D_{u}}\}\right..
\end{equation}

\textbf{Online tumor synthesis}. Specifically, we synthesize tumors in an online manner during segmentation training, which means we do not need to generate and save the synthetic tumors as offline datasets. There are two merits behind the online generation: (1) offline synthetic datasets may introduce problems about misinformation propagation of patients~\cite{chen2021synthetic}; (2) online generation enables more diverse synthesis, enabling us to synthesize a large number of tumors for segmentation training. 

\subsection{Datasets and Implementation Details}
\label{sec_method_implementation}

\textbf{Datasets collection and pre-processing}. Our proposed FreeTumor excels in leveraging large-scale data for tumor synthesis and segmentation. Thus, in this study, we first create a large-scale dataset with 161,130 publicly available CT volumes from 33 different sources, as shown in \textbf{Extended Data Table~\ref{table_dataset}}. 

As described in \textbf{Section~\ref{sec_method_synthesis}}, our initial step involves simulating tumor positions within their corresponding organ regions, \emph{e.g.}, liver tumors on livers, pancreas tumors on pancreases. Consequently, generating the organ labels becomes essential. While a few of the datasets already include organ labels, the others still lack organ labels. To address this, we first utilize a robust organ segmentation model VoCo~\cite{voco, voco-v1} to generate liver, pancreas, and kidney labels for the abdomen CT datasets. For lung organs, we employ Lungmask~\cite{lungmask} to generate lung labels for chest CT datasets. This approach enables us to leverage the entirety of 161,130 CT volumes for tumor synthesis and segmentation training. 
Note that we only utilize the generated organ labels to simulate approximate tumor positions. Therefore, these organ labels do not need to be perfectly precise for the scope of this study.

Among our curated datasets, some of them contain abdomen regions, some of them contain chest regions, and a few of them contain both abdomen and chest regions. Specifically, for the training of liver, pancreas, and kidney tumors, we utilize 19,571 abdomen CT volumes for training. For lung tumors and COVID-19, we utilize 141,784 chest CT volumes for training.

\textbf{Implementation details.} In this study, instead of developing new network architectures, we mainly focus on advancing tumor segmentation from a data-driven aspect. Thus, we simply adopt the SwinUNETR~\cite{swin} as the tumor segmentation model. We use SwinUNETR~\cite{swin} for two reasons: (1) It achieves competitive results among the baseline tumor segmentation methods~\cite{UNET,transunet,unetr,nnUNet,swin}. (2) Previous tumor synthesis methods~\cite{Syntumor,Difftumor} and CT foundation models~\cite{SwinSSL,voco-v1} also adopt SwinUNETR~\cite{swin} as backbones.

We use Pytorch~\cite{pytorch}, MONAI~\cite{monai}, and nnUNet~\cite{nnUNet} framework to conduct all the experiments. The synthesis training of FreeTumor is conducted on 8 $\times$ NVIDIA H800 (80G) GPUs. The tumor segmentation training is conducted on 1 $\times$ NVIDIA H800 (80G) GPU. More implementation details are presented in \textbf{Extended Data Table~\ref{table_preprocess}}.

\textbf{Evaluation metrics.} For the Visual Turing Test in clinician evaluation, we report the sensitivity, specificity, and accuracy results to measure the radiologists' ability of identifying synthetic tumors. Sensitivity (\%) and specificity (\%) are calculated as:
\begin{equation}\label{eqn_sensi_speci}
    \text{sensitivity} = \frac{\text{TP}}{\text{TP} + \text{FN}}, ~\text{specificity} = \frac{\text{TN}}{\text{TN} + \text{FP}},
\end{equation}
where TP (True Positive) denotes truly identifying the synthetic tumors, TN (True Negative) denotes truly identifying the real tumors, FP (False Positive) denotes falsely recognizing real tumors as synthetic tumors, and FN (False Negative) denotes falsely recognizing synthetic tumors as real tumors. The accuracy (\%) is calculated as:
\begin{equation}\label{eqn_acc}
    \text{accuracy} = \frac{\text{TP} + \text{TN}}{\text{TP} + \text{TN} + \text{FP} + \text{FN}}.
\end{equation}

For tumor segmentation, the standard Dice scores (\%) is employed to evaluate the performances. Dice scores is calculated as:
\begin{equation}\label{eqn_dsc}
    \text{Dice}(Pre, Gro) = \frac{2 |Pre \cap Gro|}{|Pre| + |Gro|},
\end{equation}
where $Pre$ denotes the segmentation predictions, $Gro$ is the ground truth of tumor labels. 

For tumor detection, detected tumors are identified when segmentation predictions overlap with the ground truth labels~\cite{cao2023large,fitzgerald2022future,singhi2019early,pereira2020early}. We use F1-Score, sensitivity, and specificity to measure the performances of tumor detection, where F1-Score is formulate as:
\begin{equation}\label{eqn_f1_score}
    \text{F1-Score} = \frac{2 \times \text{Precision} \times \text{Recall}}{\text{Precision} + \text{Recall}},
\end{equation}
where Precision and Recall are formulated as:
\begin{equation}\label{eqn_pre_recall}
     \text{Precision} = \frac{\text{TP}}{\text{TP} + \text{FP}}, ~\text{Recall} = \frac{\text{TP}}{\text{TP} + \text{FN}}.
\end{equation}
Here, the ``positive'' class is defined as detecting a tumor within a CT volume. 
    
\section{Data availability}\label{sec_data_avail}

This study incorporates a total of 33 public datasets from different sources, encompassing 161,130 publicly available CT volumes. All these datasets are publicly available for research. For detailed information about the data used
in this project, please refer to \textbf{Extended Data Table~\ref{table_dataset}}. 

\section{Code availability}\label{sec6}

The codes, datasets, and models of FreeTumor are available at GitHub
(\href{https://github.com/Luffy03/FreeTumor}{https://github.com/Luffy03/FreeTumor}).

\section{Author contributions}\label{sec8}

L.W. designed the framework and conducted the experiments. Y.Z., L.L., X.W., and P.R. provided suggestions on the framework and experiments. J.Z., S.H., J.M., and X.N. contributed to the data acquisition and downstream task evaluation. X.Z, M.W, Y.W, X.D, and V.V. contributed to the clinician evaluation of tumor synthesis and analyzed the results of tumor recognition. All authors contributed to the drafting and revising of the manuscript. H.C. conceived and supervised the work.


\section*{Declaration}

The authors have no conflicts of interest to declare.

\section*{Ethics declaration}\label{sec7}

This project has been reviewed and approved by the Human and Artefacts Research Ethics Committee (HAREC). The protocol number is HREP-2024-0429.

\section*{Acknowledgements}
This work was supported by the Hong Kong Innovation and Technology Commission (Project No. MHP/002/22, GHP/006/22GD and ITCPD/17-9), HKUST (Project No. FS111), and the Research Grants Council of the Hong Kong Special Administrative Region, China (Project Reference Number: T45-401/22-N). 
We also thank the support of HKUST SuperPod for providing the GPU platform for model training. 
We express our sincere gratitude to the radiologists who contributed to the clinician evaluation, including Shisi Li, Dexuan Chen, Lingling Yang, Yu Wang, Riyu Han, Lin Liu, Kanrong Yang, Rui Zhang, Guangzi Shi, and Qiang Ye. We greatly appreciate their dedicated efforts.


\bibliography{sn-bibliography}

\newpage

\begin{appendices}

\section{Extended Data}\label{secA1}

\begin{figure*}[!h]
	\centering
	\includegraphics[width=1\linewidth]{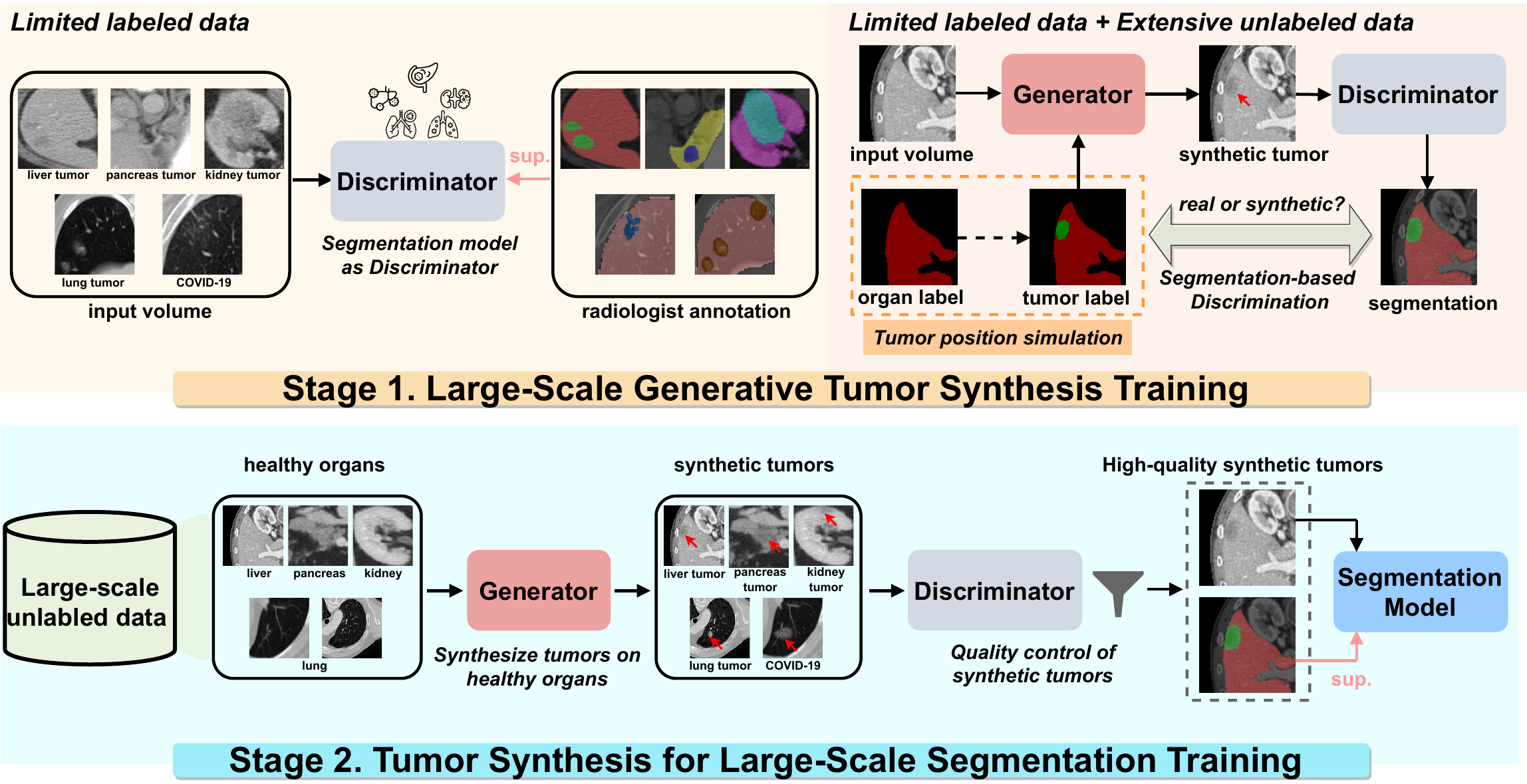}
	\caption{\textbf{The framework of FreeTumor}, including two stages: \textbf{(1)} \emph{Large-Scale Generative Tumor Synthesis Training}. We first leverage labeled data to train a baseline segmentation model as the discriminator of the tumor synthesis model. Second, we leverage both labeled and unlabeled data to train the tumor synthesis model. Specifically, we train the generator to synthesize tumors on health organs while the discriminator is utilized to discriminate the reality of synthetic tumors. \textbf{(2)} \emph{Tumor Synthesis for Large-Scale Segmentation Training}. We employ the generator to synthesize tumors on healthy organs for augmenting segmentation training datasets, while the discriminator is employed for quality control of synthetic tumors. Specifically, \emph{Sup.} denotes employ labels for supervision.}
	\label{fig_frame}
\end{figure*}

\clearpage

\renewcommand{\arraystretch}{1.12}
\begin{table*}[!h]
    \setlength{\abovecaptionskip}{0pt}
    \setlength{\belowcaptionskip}{-0.2em}
    \centering
    \footnotesize
    \caption{The sensitivity (sen.) ($\%$) and specificity (spe.) ($\%$) results of Visual Turing Test. The radiologists are divided into three groups, \emph{i.e.}, Junior (Jun), Mid-level (Mid), and Senior (Sen). The best results are \textbf{bolded} and the standard deviations (std) are reported.}
    \vspace{.03in}
    \begin{threeparttable}
        \begin{tabular}{p{1cm}p{0.7cm}p{0.7cm}|p{0.7cm}p{0.7cm}|p{0.7cm}p{0.7cm}|p{0.7cm}p{0.7cm}|p{0.6cm}p{0.6cm}}
            \toprule[1.2pt]
             & \multicolumn{2}{c}{\textbf{Liver}} &\multicolumn{2}{c}{\textbf{Panc.}} &\multicolumn{2}{c}{\textbf{Kid.}} &\multicolumn{2}{c}{\textbf{Lung}} &\multicolumn{2}{c}{\textbf{COVID}}\\
            &sen. &spe. &sen. &spe. &sen. &spe. &sen. &spe. &sen. &spe.\\
            \hline
            \textbf{Jun-1} &11.1 &55.6 &11.1 &88.9 &22.2 &77.8 &0.0 &88.9 &0.0 &88.9\\
            
            \textbf{Jun-2} &55.6 &55.6 &22.2 &55.6 &22.2 &55.6 &77.8 &77.8 &77.8 &55.6\\

            \textbf{Jun-3} &55.6 &66.7 &11.1 &88.9 &77.8 &77.8 &55.6 &88.9 &55.6 &44.4\\

            \textbf{Jun-4} &55.6 &55.6 &22.2 &77.8 &77.8 &66.7 &66.7 &77.8 &77.8 &66.7\\

            \textbf{Jun-5} &66.7 &66.7 &22.2 &66.7 &77.8 &77.8 &66.7 &77.8 &77.8 &77.8\\

            \textbf{Jun-6} &11.1 &77.8 &0.0 &88.9 &0.0 &88.9 &33.3 &55.6 &33.3 &66.7\\
            
            \hline

            \textbf{Mid-1} &55.6 &55.6 &11.1 &77.8 &22.2 &66.7 &0.0 &66.7 &11.1 &88.9\\
            
            \textbf{Mid-2} &44.4 &44.4 &22.2 &66.7 &22.2 &77.8 &77.8 &33.3 &55.6 &77.8\\

            \textbf{Mid-3} &55.6 &88.9 &\textbf{77.8} &\textbf{66.7} &77.8 &77.8 &77.8 &77.8 &55.6 &66.7\\

            \textbf{Mid-4} &66.7 &66.7 &44.4 &66.7 &55.6 &66.7 &\textbf{88.9} &\textbf{88.9} &88.9 &66.7\\
            
            \hline
            \textbf{Sen-1} &66.7 &55.6 &55.6 &55.6 &100.0 &66.7 &66.7 &88.9 &66.7 &22.2\\

            \textbf{Sen-2} &\textbf{77.8} &\textbf{88.9} &33.3 &77.8 &55.6 &66.7 &55.6 &88.9 &66.7 &66.7\\
            
            \textbf{Sen-3} &77.8 &77.8 &66.7 &66.7 &\textbf{100.0} &\textbf{77.8} &88.9 &44.4 &\textbf{88.9} &\textbf{88.9}\\

            \hline
            \rowcolor{mygray}
            \textbf{Average} & 53.9 & 65.8 & 30.8 & 72.7 & 54.7 & 72.7 &58.1 & 73.5 & 58.1 & 67.5 \\

            \rowcolor{mygray}
            \textbf{std} & 20.3 & 15.6 & 23.5 & 10.4 & 33.2 & 7.7 &26.5 & 18.3 & 26.5 & 18.3 \\

            \toprule[1.2pt]
        \end{tabular}
    \end{threeparttable}
    \label{table_turing_test_sen}
\end{table*}

\vspace{-.2in}
\begin{table*}[!h]
    \setlength{\abovecaptionskip}{0pt}
    \setlength{\belowcaptionskip}{-0.2em}
    \centering
    \footnotesize
    \caption{The accuracy ($\%$) results of Visual Turing Test.}
     
    \vspace{.03in}
    \begin{threeparttable}
        \begin{tabular}{lccccc}
            \toprule[1.2pt]
             &\textbf{Liver} &\textbf{Panc.} &\textbf{Kid.} &\textbf{Lung} &\textbf{COVID}\\
            \hline
            \textbf{Jun-1} &33.3 &50.0 &50.0 &44.4 &44.4\\
            
            \textbf{Jun-2} &55.6 &38.9 &38.9 &77.8 &66.7\\

            \textbf{Jun-3} &61.2 &50.0 &77.8 &72.2 &50.0\\

            \textbf{Jun-4} &55.6 &50.0 &72.2 &72.2 &72.2\\

            \textbf{Jun-5} &66.7 &44.4 &77.8 &72.2 &77.8\\

            \textbf{Jun-6} &44.4 &44.4 &44.4 &44.4 &50.0\\
            
            \hline

            \textbf{Mid-1} &55.6 &44.4 &44.4 &33.3 &50.0\\
            
            \textbf{Mid-2} &44.4 &44.4 &50.0 &55.6 &66.7\\

            \textbf{Mid-3} &72.2 &\textbf{72.2} &77.8 &77.8 &61.1\\

            \textbf{Mid-4} &66.7 &55.6 &61.1 &\textbf{88.9} &77.8\\
            
            \hline
            \textbf{Sen-1} &61.1 &55.6 &83.3 &77.8 &44.4\\

            \textbf{Sen-2} &\textbf{83.3} &55.6 &61.1 &72.2 &66.7\\

            \textbf{Sen-3} &77.8 &66.7 &\textbf{88.9} &66.7 &\textbf{88.9}\\
            
            \hline
            \rowcolor{mygray}
            \textbf{Average} & $59.8$ & $51.7$ & $63.7$ & $65.8$ & $62.8$ \\

            \rowcolor{mygray}
            \textbf{std} & 15.5 & 9.1 & 15.9 & 14.9 & 13.7 \\

            \toprule[1.2pt]
        \end{tabular}
    \end{threeparttable}
    \label{table_turing_test_acc}
\end{table*}

\begin{table*}[!h]
    \setlength{\abovecaptionskip}{0pt}
    \setlength{\belowcaptionskip}{-0.2em}
    \centering
    \footnotesize
    \caption{Average (Avg.) sensitivity ($\%$), specificity ($\%$), and accuracy ($\%$) of Visual Turing Test for junior, mid-level, and senior radiologists, respectively.}
    \vspace{.03in}
    \begin{threeparttable}
        \begin{tabular}{c|cc|c}
            \toprule[1.2pt]
             &\textbf{Avg. sensitivity} &\textbf{Avg. specificity} &\textbf{Avg. accuracy}\\
            \hline
            6 Junior radiologists &${41.5}$ &${72.2}$ &${56.6}$\\
            4 Mid-level radiologists &${50.5}$ &${69.4}$ &${60.0}$\\
            3 Senior radiologists &${71.8}$ &${68.3}$ &${70.0}$\\
            \hline
            Total &${51.1}$ &${70.4}$ &${60.8}$\\
            \toprule[1.2pt]
        \end{tabular}
    \end{threeparttable}
    \label{table_average_turing}
\end{table*}

\clearpage

\renewcommand{\arraystretch}{1.5}
\begin{table*}[!h]
    \setlength{\abovecaptionskip}{0pt}
    \setlength{\belowcaptionskip}{-0.2em}
    \centering
    \footnotesize
    \caption{\textbf{Quantitative Fréchet Inception Distance (FID)~\cite{fid} results of synthetic tumors}. We crop the tumor regions and calculate the FID between real and synthetic tumors. Note that \textbf{lower FID represents higher synthesis quality}. We report the results of SynTumor~\cite{Syntumor} and DiffTumor~\cite{Difftumor} for comparisons. SynTumor is only applicable to liver tumors, and DiffTumor is not applicable to lung tumors and COVID-19. We report the standard deviation via five times of experiments.}
    \vspace{.03in}
    \begin{threeparttable}
        \begin{tabular}{cccccc}
            \toprule[1.2pt]
             &\textbf{Liver} &\textbf{Panc.} &\textbf{Kid.} &\textbf{Lung} &\textbf{COVID}\\
            \hline
            SynTumor~\cite{Syntumor} &${45.1}_{{\pm}6.5}$ &\XSolidBrush &\XSolidBrush &\XSolidBrush &\XSolidBrush\\
            DiffTumor~\cite{Difftumor} &${33.7}_{{\pm}2.7}$ &${144.2}_{{\pm}8.6}$ &${16.4}_{{\pm}1.0}$ &\XSolidBrush &\XSolidBrush\\
            FreeTumor &${23.5}_{{\pm}1.2}$ &${72.3}_{{\pm}4.8}$ &${15.7}_{{\pm}0.8}$ &${23.2}_{{\pm}1.1}$ &${29.1}_{{\pm}1.5}$\\
            \toprule[1.2pt]
        \end{tabular}
    \end{threeparttable}
    \label{table_fid}
\end{table*}

\begin{figure*}[!h]
	\centering
	\includegraphics[width=1\linewidth]{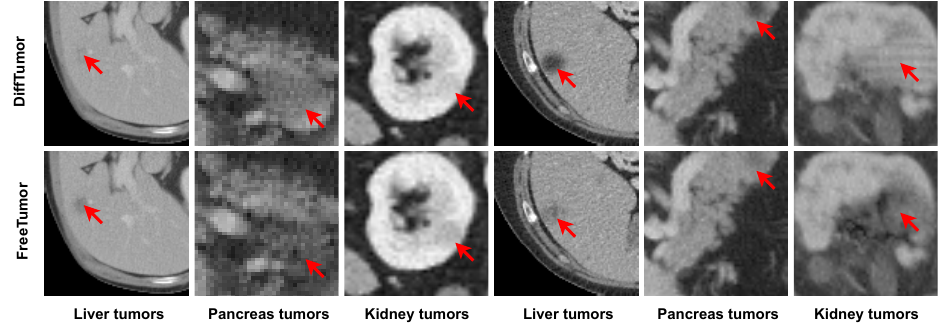}
	\caption{Qualitative comparisons with DiffTumor. We synthesize tumors at the same positions of CT volumes for comparisons.}
	\label{fig_vsDiff}
\end{figure*}
\vspace{-8in}

\begin{figure*}[!h]
	\centering
	\includegraphics[width=1\linewidth]{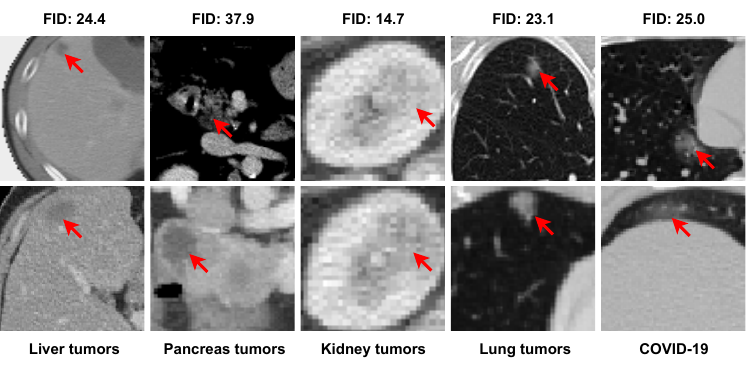}
	\caption{We present some failure cases of FreeTumor. We observe that FID is not very reliable in reflecting tumor synthesis quality. The presented synthetic tumors are with low FIDs but exhibit unrealistic features for radiologists. To this end, we highlight the clinician evaluation results in the main paper, which is more convincing.}
	\label{fig_FID}
\end{figure*}

\begin{figure*}[!h]
	\centering
	\includegraphics[width=1\linewidth]{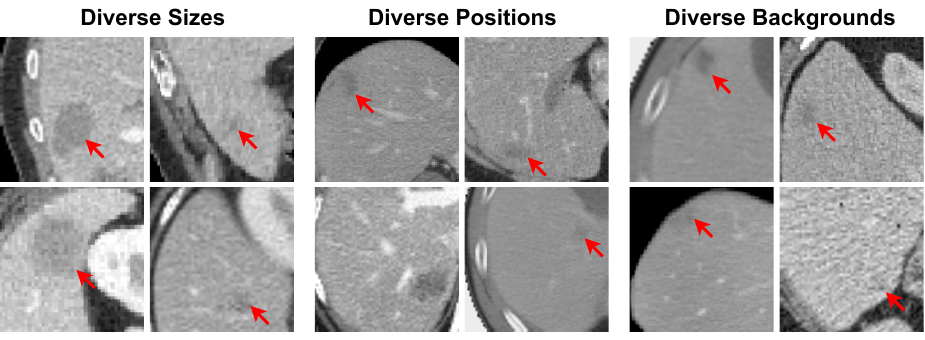}
	\caption{\textbf{Diversity of synthetic tumors}. FreeTumor can synthesize tumors of varying sizes and positions. In addition, by aggregating large-scale CT datasets with varying image characteristics, FreeTumor also provides a wide range of backgrounds for synthetic tumors.}
	\label{fig_diversity}
\end{figure*}
\clearpage

\begin{figure*}
	\centering
	\includegraphics[width=0.9\linewidth]{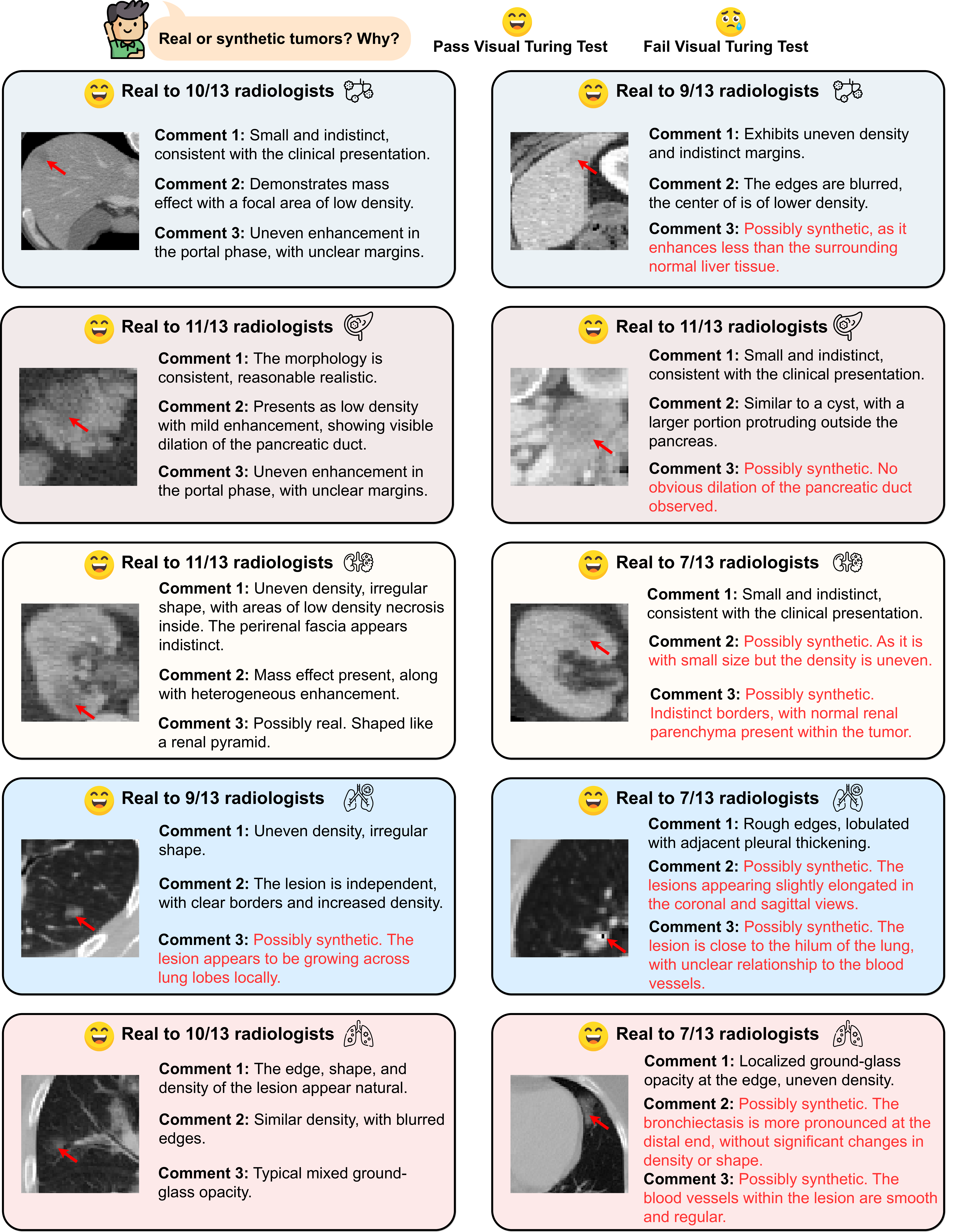}

	\caption{\textbf{Case studies}. From the first to the fifth row, we present the synthetic results of liver tumors, pancreas tumors, kidney tumors, lung tumors, and COVID-19, respectively. The synthetic tumors/lesions are highlighted by \textcolor{red}{red arrows}. For each case, we select three representative comments from different levels of radiologists. Specifically, for radiologists who successfully identified the synthetic tumors, we highlight their comments with \textcolor{red}{red color}. Additionally, we present the number of radiologists who mistakenly identified the synthetic case as real, \emph{e.g.}, in the first box, ``Real to 10/13 radiologists'' means 10 of 13 radiologists misclassified a synthetic tumor as a real one. More case studies with failure cases are presented in \textbf{Extended Data Figure~\ref{fig_Case_study2}}.}
	\label{fig_Case_study1}
\end{figure*}
\clearpage

\begin{figure*}
	\centering
	\includegraphics[width=0.95\linewidth]{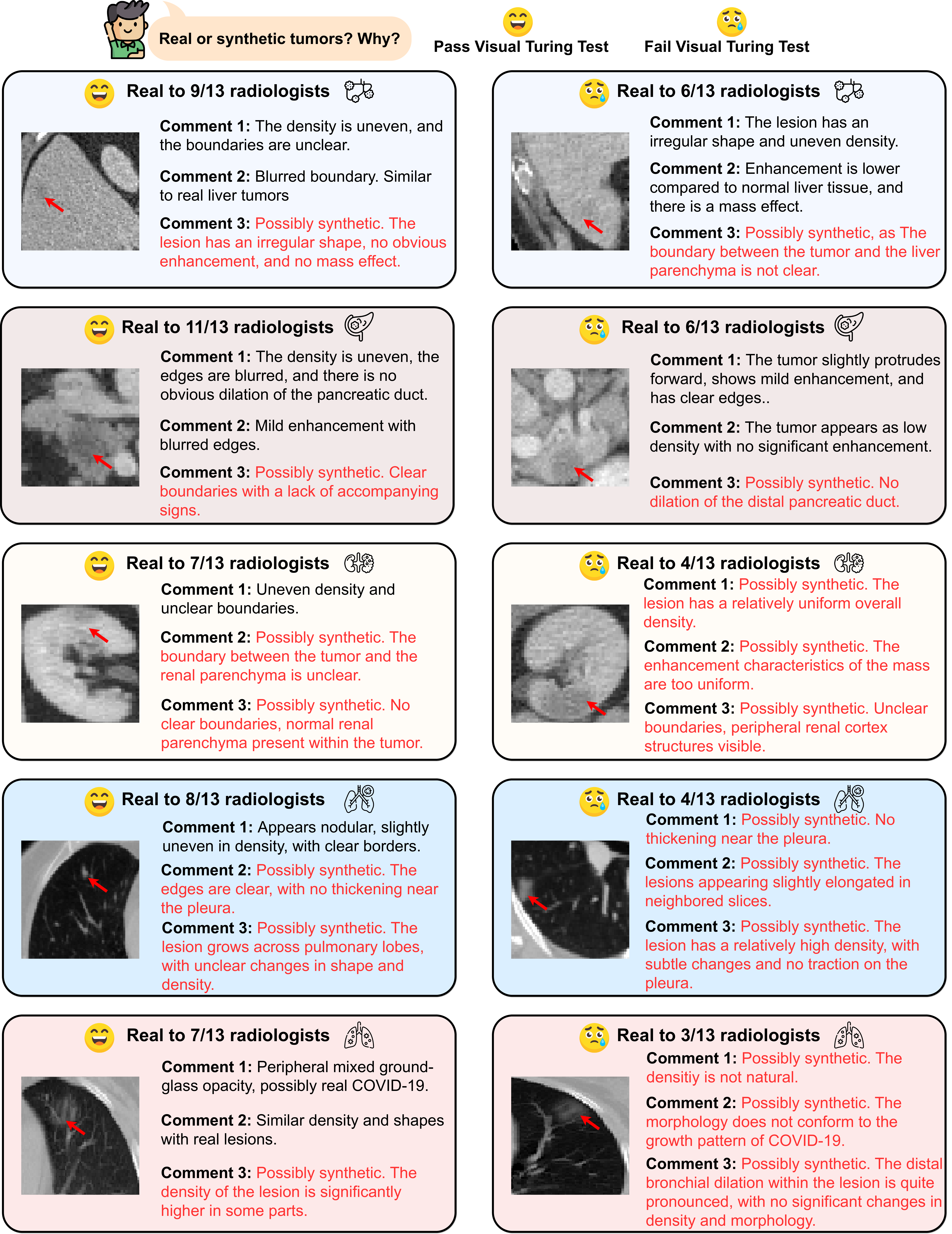}
	\caption{Additional case studies of synthetic tumors are provided. We present more failure cases as a complement to \textbf{Figure~\ref{fig_Case_study1}}.}
	\label{fig_Case_study2}
\end{figure*}

\clearpage

\begin{table*}
    \setlength{\abovecaptionskip}{0pt}
    \setlength{\belowcaptionskip}{-0.2em}
    \centering
    \footnotesize
              \caption{\textbf{Quantitative results of tumor segmentation}. Given the limited scale of public datasets, we conducted 5-fold evaluations on 12 widely-used tumor datasets. For the confidence range, we report the results of the best and worst validation folds. The best results are \textbf{bolded} while the second best results are \underline{underlined}.}
    \vspace{.03in}
    \begin{threeparttable}
        \begin{tabular}{p{2cm}p{1.3cm}p{1.3cm}p{1.3cm}|p{1.3cm}p{1.3cm}p{1.3cm}}
            \toprule[1.2pt]
            \multirow{2}{*}{\textbf{Method}} & \multicolumn{3}{c}{\textbf{Liver Tumor}} & \multicolumn{3}{c}{\textbf{Pancreas Tumor}} \\
            &LiTS~\cite{LITS} &HCC.~\cite{HCC-TACE-Seg} &IR.~\cite{3D-IRCADb} &MSD07~\cite{msd} &PANO.~\cite{PANORAMA} &QUBIQ~\cite{Qubiq} \\
            \hline
            \multirow{2}{*}{UNet~\cite{UNET}} &${49.2}$ &${62.3}$ &${37.8}$ &${44.2}$ &${39.3}$ &${43.1}$ \\
            
            &$(43.7,52.4)$ &$(57.0,65.8)$ &$(24.3,48.9)$ &$(35.8,50.1)$ &$(31.8,49.5)$ &$(34.5,59.2)$ \\
            
             \multirow{2}{*}{TransUNet~\cite{transunet}} &${53.2}$ &${69.8}$ &${44.3}$ &${47.7}$ &${33.4}$ &${49.1}$ \\
             
             &$(48.5,60.0)$ &$(65.1,74.1)$ &$(28.9,49.4)$ &$(39.5,54.9)$ &$(29.1,39.6)$ &$(38.9,59.6)$ \\
            
             \multirow{2}{*}{UNETR~\cite{unetr}} &${51.3}$ &${65.2}$ &${37.3}$ &${51.5}$ &${40.4}$ &${47.9}$ \\
             
             &$(42.4,55.7)$ &$(59.6,72.4)$ &$(31.9,43.1)$ &$(43.2,56.1)$ &$(33.6,47.1)$ &$(38.3,58.0)$ \\
            
             \multirow{2}{*}{nnUNet~\cite{nnUNet}} &\underline{${60.3}$} &${71.2}$ &${42.1}$ &\underline{${52.7}$} &${46.4}$ &\underline{${52.2}$} \\
             
             &$(52.0,64.5)$ &$(67.0,75.1)$ &$(31.7,50.8)$ &$(49.5,58.5)$ &$(41.1,50.0)$ &$(42.6,63.8)$ \\
            
             \multirow{2}{*}{SwinUNETR~\cite{swin}} &${58.6}$ &\underline{${71.3}$} &\underline{${48.2}$} &${52.6}$ &\underline{${47.1}$} &${51.8}$ \\
             
             &$(54.3,65.1)$ &$(67.2,75.8)$ &$(40.6,60.5)$ &$(45.2,60.8)$ &$(42.1,52.8)$ &$(45.2,63.2)$ \\
            
             \hline
             \multirow{2}{*}{\textbf{FreeTumor}} &$\mathbf{{65.5}}$ &$\mathbf{{79.9}}$ &$\mathbf{{64.3}}$ &$\mathbf{{58.6}}$ &$\mathbf{{50.2}}$ &$\mathbf{{59.1}}$ \\
             
             &$(61.6,71.4)$ &$(75.3,84.2)$ &$(59.0,70.0)$ &$(51.9,64.6)$ &$(45.7,56.4)$ &$(51.5,68.7)$ \\
         
            \toprule[1.2pt]
            \multirow{2}{*}{\textbf{Method}} & \multicolumn{3}{c}{\textbf{Kidney Tumor}} & \multicolumn{3}{c}{\textbf{Lung Tumor \& COVID-19}} \\
            &KiTS21~\cite{kits19} &KiTS23~\cite{kits} &KIPA~\cite{KIPA} &MSD06~\cite{msd} &RIDER~\cite{RIDER-LungCT-Seg} &CV19~\cite{COVID} \\
            \hline
             \multirow{2}{*}{UNet~\cite{UNET}} &${65.2}$ &${62.6}$ &${70.8}$ &${44.9}$ &${54.0}$ &${55.3}$ \\
             
             &$(60.6,70.1)$ &$(57.7,65.8)$ &$(62.0,80.6)$ &$(41.2,50.5)$ &$(49.1,60.6)$ &$(47.8,60.6)$ \\
            
             \multirow{2}{*}{TransUNet~\cite{transunet}} &${71.6}$ &${67.3}$ &${76.0}$ &${47.0}$ &${57.3}$ &${60.9}$\\
             
             &$(60.6,75.4)$ &$(60.7,72.1)$ &$(61.4,84.9)$ &$(41.3,54.9)$ &$(49.1,68.4)$ &$(52.3,70.3)$ \\
            
             \multirow{2}{*}{UNETR~\cite{unetr}} &${69.6}$ &${70.3}$ &${76.3}$ &${54.0}$ &${61.3}$ &${62.3}$\\
             
             &$(67.9,73.3)$ &$(62.7,74.6)$ &$(68.0,86.3)$ &$(49.4,57.2)$ &$(56.7,68.4)$ &$(47.8,70.4)$ \\
            
             \multirow{2}{*}{nnUNet~\cite{nnUNet}} &${72.4}$ &\underline{${71.9}$} &${77.4}$ &\underline{${58.9}$} &\underline{${62.8}$} &\underline{${63.5}$}\\
             
             &$(65.9,78.4)$ &$(65.4,79.8)$ &$(71.1,85.8)$ &$(50.8,66.9)$ &$(57.1,70.2)$ &$(56.5,72.6)$ \\
            
             \multirow{2}{*}{SwinUNETR~\cite{swin}} &\underline{${72.5}$} &${71.6}$ &\underline{${77.5}$} &${58.7}$ &${62.5}$ &${63.3}$\\
             
             &$(68.0,79.0)$ &$(63.8,82.6)$ &$(73.9,83.7)$ &$(50.2,68.7)$ &$(57.2,70.3)$ &$(54.9,72.8)$ \\
           
             \hline
             \multirow{2}{*}{\textbf{FreeTumor}} &$\mathbf{{74.5}}$ &$\mathbf{{75.3}}$ &$\mathbf{{83.3}}$ &$\mathbf{{65.8}}$ &$\mathbf{{67.6}}$ &$\mathbf{{71.2}}$ \\
             
             &$(69.3,80.5)$ &$(70.2,84.4)$ &$(79.6,87.4)$ &$(60.4,70.7)$ &$(60.3,74.9)$ &$(64.8,77.7)$ \\

            \toprule[1.2pt]
        \end{tabular}
    \end{threeparttable}
    \label{table_results}
\end{table*}

\clearpage

\begin{table*}
    \setlength{\abovecaptionskip}{0pt}
    \setlength{\belowcaptionskip}{-0.2em}
    \centering
    \footnotesize
      \caption{\textbf{Comparison with baseline tumor segmentation models.} We report the Dice score ($\%$) results on 12 datasets. Copy-Paste~\cite{cutmix}, SynTumor~\cite{Syntumor}, DiffTumor~\cite{Difftumor}, and FreeTumor all adopt SwinUNETR~\cite{swin} as the segmentation model. Copy-Paste is based on Cutmix~\cite{cutmix}, which simply cuts the real tumors to the healthy regions. SynTumor is only applicable to liver tumors, and DiffTumor is not applicable to lung tumors and COVID-19. We conduct 5-fold evaluations and report the results of the best and worst validation folds for the confidence range.}
    \vspace{.03in}
    \begin{threeparttable}
        \begin{tabular}{p{2cm}p{1.3cm}p{1.3cm}p{1.3cm}|p{1.3cm}p{1.3cm}p{1.3cm}}
            \toprule[1.2pt]
            \multirow{2}{*}{\textbf{Method}} & \multicolumn{3}{c}{\textbf{Liver Tumor}} & \multicolumn{3}{c}{\textbf{Pancreas Tumor}} \\
            &LiTS~\cite{LITS} &HCC.~\cite{HCC-TACE-Seg} &IR.~\cite{3D-IRCADb} &MSD07~\cite{msd} &PANO.~\cite{PANORAMA} &QUBIQ~\cite{Qubiq} \\
            \hline
            \multirow{2}{*}{Copy-Paste~\cite{cutmix}} &${22.5}$ &${22.3}$ &${17.8}$ &${10.2}$ &${8.3}$ &${10.1}$ \\
            
            &$(13.7,27.4)$ &$(17.0,25.8)$ &$(14.3,20.8)$ &$(8.8,11.1)$ &$(7.8,9.0)$ &$(8.5,12.2)$ \\
            
             \multirow{2}{*}{SynTumor~\cite{Syntumor}} &${60.2}$ &${72.6}$ &${46.2}$ &\XSolidBrush &\XSolidBrush &\XSolidBrush \\
             
             &$(56.2,66.4)$ &$(67.2,77.4)$ &$(41.0,51.9)$ &\XSolidBrush &\XSolidBrush &\XSolidBrush \\
            
             \multirow{2}{*}{DiffTumor~\cite{Difftumor}} &${62.3}$ &${75.8}$ &${52.4}$ &${54.5}$ &${41.9}$ &${48.0}$ \\
             
             &$(58.0,68.3)$ &$(70.1,79.3)$ &$(31.9,43.1)$ &$(48.6,60.9)$ &$(37.9,48.6)$ &$(41.1,57.8)$ \\
            
             \hline
             \multirow{2}{*}{\textbf{FreeTumor}} &$\mathbf{{65.5}}$ &$\mathbf{{79.9}}$ &$\mathbf{{64.3}}$ &$\mathbf{{58.6}}$ &$\mathbf{{50.2}}$ &$\mathbf{{59.1}}$ \\
             
             &$(61.6,71.4)$ &$(75.3,84.2)$ &$(59.0,70.0)$ &$(51.9,64.6)$ &$(45.7,56.4)$ &$(51.5,68.7)$ \\
         
            \toprule[1.2pt]
            \multirow{2}{*}{\textbf{Method}} & \multicolumn{3}{c}{\textbf{Kidney Tumor}} & \multicolumn{3}{c}{\textbf{Lung Tumor \& COVID-19}} \\
            &KiTS21~\cite{kits19} &KiTS23~\cite{kits} &KIPA~\cite{KIPA} &MSD06~\cite{msd} &RIDER~\cite{RIDER-LungCT-Seg} &CV19~\cite{COVID} \\
            \hline
            
            \multirow{2}{*}{Copy-Paste~\cite{cutmix}} &${37.6}$ &${40.1}$ &${34.8}$ &${44.2}$ &${38.7}$ &${33.0}$ \\
            
            &$(33.6,44.4)$ &$(35.0,46.2)$ &$(30.3,41.9)$ &$(37.7,48.1)$ &$(34.9,47.0)$ &$(24.4,39.5)$ \\
            
             \multirow{2}{*}{SynTumor~\cite{Syntumor}} &\XSolidBrush &\XSolidBrush &\XSolidBrush &\XSolidBrush &\XSolidBrush &\XSolidBrush \\
             
             &\XSolidBrush &\XSolidBrush &\XSolidBrush &\XSolidBrush &\XSolidBrush &\XSolidBrush \\
            
             \multirow{2}{*}{DiffTumor~\cite{Difftumor}} &${70.1}$ &${70.4}$ &${58.9}$ &\XSolidBrush &\XSolidBrush &\XSolidBrush \\
             
             &$(65.3,75.8)$ &$(64.0,79.9)$ &$(55.5,63.6)$ &\XSolidBrush &\XSolidBrush &\XSolidBrush \\
           
             \hline
             \multirow{2}{*}{\textbf{FreeTumor}} &$\mathbf{{74.5}}$ &$\mathbf{{75.3}}$ &$\mathbf{{83.3}}$ &$\mathbf{{65.8}}$ &$\mathbf{{67.6}}$ &$\mathbf{{71.2}}$ \\
             
             &$(69.3,80.5)$ &$(70.2,84.4)$ &$(79.6,87.4)$ &$(60.4,70.7)$ &$(60.3,74.9)$ &$(64.8,77.7)$ \\

            \toprule[1.2pt]
        \end{tabular}
    \end{threeparttable}
    \label{table_compare_synthesis}
\end{table*}

\clearpage

\begin{table*}
    \setlength{\abovecaptionskip}{0pt}
    \setlength{\belowcaptionskip}{-0.2em}
    \centering
    \footnotesize
      \caption{\textbf{Comparison with Self-Supervised Learning (SSL) CT foundation models}. SwinUNETR~\cite{swin} is adopted as the segmentation model. We report the Dice score ($\%$) results on 12 public datasets. We conduct 5-fold evaluations and report the results of the best and worst validation folds for the confidence range.}
    \vspace{.03in}
    \begin{threeparttable}
        \begin{tabular}{p{2cm}p{1.3cm}p{1.3cm}p{1.3cm}|p{1.3cm}p{1.3cm}p{1.3cm}}
            \toprule[1.2pt]
            \multirow{2}{*}{\textbf{Method}} & \multicolumn{3}{c}{\textbf{Liver Tumor}} & \multicolumn{3}{c}{\textbf{Pancreas Tumor}} \\
            &LiTS~\cite{LITS} &HCC.~\cite{HCC-TACE-Seg} &IR.~\cite{3D-IRCADb} &MSD07~\cite{msd} &PANO.~\cite{PANORAMA} &QUBIQ~\cite{Qubiq} \\
            \hline
            \multirow{2}{*}{MAE3D~\cite{MAE3D,MAE}} &${56.8}$ &${66.2}$ &${35.5}$ &${43.7}$ &${35.8}$ &${49.0}$ \\
            
            &$(52.7,62.5)$ &$(57.0,65.8)$ &$(27.1,40.8)$ &$(36.6,51.6)$ &$(30.8,44.2)$ &$(41.6,59.4)$ \\
            
             \multirow{2}{*}{SwinSSL~\cite{SwinSSL}} &${58.3}$ &${69.7}$ &${44.4}$ &${50.6}$ &${45.7}$ &${50.4}$ \\
             
             &$(54.0,64.3)$ &$(67.0,72.1)$ &$(39.9,49.1)$ &$(43.1,55.8)$ &$(41.3,50.3)$ &$(42.2,61.0)$ \\
            
             \multirow{2}{*}{VoCo~\cite{voco-v1}} &${60.5}$ &${71.2}$ &${50.7}$ &${51.8}$ &${47.6}$ &${52.9}$ \\
             
             &$(56.3,65.4)$ &$(67.3,77.4)$ &$(45.8,56.2)$ &$(45.6,58.4)$ &$(42.5,52.6)$ &$(45.6,62.8)$ \\
            
             \hline
             \multirow{2}{*}{\textbf{FreeTumor}} &$\mathbf{{65.5}}$ &$\mathbf{{79.9}}$ &$\mathbf{{64.3}}$ &$\mathbf{{58.6}}$ &$\mathbf{{50.2}}$ &$\mathbf{{59.1}}$ \\
             
             &$(61.6,71.4)$ &$(75.3,84.2)$ &$(59.0,70.0)$ &$(51.9,64.6)$ &$(45.7,56.4)$ &$(51.5,68.7)$ \\
         
            \toprule[1.2pt]
            \multirow{2}{*}{\textbf{Method}} & \multicolumn{3}{c}{\textbf{Kidney Tumor}} & \multicolumn{3}{c}{\textbf{Lung Tumor \& COVID-19}} \\
            &KiTS21~\cite{kits19} &KiTS23~\cite{kits} &KIPA~\cite{KIPA} &MSD06~\cite{msd} &RIDER~\cite{RIDER-LungCT-Seg} &CV19~\cite{COVID} \\
            \hline
            
            \multirow{2}{*}{MAE3D~\cite{MAE3D,MAE}} &${65.1}$ &${67.2}$ &${70.9}$ &${53.8}$ &${55.7}$ &${58.0}$ \\
            
            &$(60.8,72.1)$ &$(65.2,72.8)$ &$(62.5,73.9)$ &$(50.3,58.7)$ &$(50.5,63.5)$ &$(52.1,64.9)$ \\
            
             \multirow{2}{*}{SwinSSL~\cite{SwinSSL}} &${72.1}$ &${70.6}$ &${76.3}$ &${58.6}$ &${62.1}$ &${62.1}$ \\
             
             &$(67.3, 78.2)$ &$(65.3,77.2)$ &$(73.6,81.4)$ &$(53.6, 63.4)$ &$(56.1,67.4)$ &$(57.5,68.0)$ \\
            
             \multirow{2}{*}{VoCo~\cite{voco-v1}} &${72.4}$ &${72.0}$ &${77.5}$ &${59.6}$ &${63.4}$ &${65.0}$ \\
             
             &$(67.5,77.6)$ &$(66.1,80.0)$ &$(73.7,81.8)$ &$(54.7,64.4)$ &$(58.8,69.2)$ &$(58.8,71.2)$ \\
           
             \hline
             \multirow{2}{*}{\textbf{FreeTumor}} &$\mathbf{{74.5}}$ &$\mathbf{{75.3}}$ &$\mathbf{{83.3}}$ &$\mathbf{{65.8}}$ &$\mathbf{{67.6}}$ &$\mathbf{{71.2}}$ \\
             
             &$(69.3,80.5)$ &$(70.2,84.4)$ &$(79.6,87.4)$ &$(60.4,70.7)$ &$(60.3,74.9)$ &$(64.8,77.7)$ \\

            \toprule[1.2pt]
        \end{tabular}
    \end{threeparttable}
    \label{table_compare_ssl}
\end{table*}

\clearpage

\begin{sidewaystable*}
	\setlength{\abovecaptionskip}{0.pt}
	\setlength{\belowcaptionskip}{-0.em}
	\centering
	\footnotesize
 \caption{\textbf{Out-of-domain evaluation results of tumor segmentation}. We report the Dice score ($\%$) results. We train the model on a source dataset and conduct direct inference on a target dataset without fine-tuning. The standard deviations are obtained from five times of experiments.}
\begin{threeparttable}
	\begin{tabular}{l|l|ccccc|cc|ccc|c}
		\toprule[1.2pt]
		\textbf{Source} &\textbf{Target} &UNet &TransUNet &UNETR &nnUNet &Swin. &SynTumor &DiffTumor &MAE3D &SwinSSL &VoCo &FreeTumor
        \\
        \hline
        
        LiTS &HCC. &${26.7}_{{\pm}0.9}$ &${32.3}_{{\pm}1.1}$ &${30.9}_{{\pm}0.8}$ &${40.5}_{{\pm}1.9}$ &${41.3}_{{\pm}2.1}$
        
        &${40.6}_{{\pm}0.7}$ &${50.1}_{{\pm}1.0}$ 
        
        &${36.5}_{{\pm}1.2}$ &${41.9}_{{\pm}0.9}$ &${43.8}_{{\pm}1.1}$ 
        
        &$\mathbf{{57.4}_{{\pm}0.9}}$\\

        LiTS &IRCAD &${42.1}_{{\pm}1.2}$ &${45.6}_{{\pm}0.7}$ &${42.6}_{{\pm}0.8}$ &${51.2}_{{\pm}0.8}$ &${48.7}_{{\pm}0.5}$ 
        
        &${51.2}_{{\pm}1.6}$ &${56.3}_{{\pm}0.7}$ 
        
        &${46.8}_{{\pm}0.8}$ &${50.8}_{{\pm}1.1}$ &${53.3}_{{\pm}1.5}$ 
        
        &$\mathbf{{71.6}_{{\pm}0.8}}$\\
        
        MSD07 &PANO. &${23.5}_{{\pm}1.1}$ &${30.8}_{{\pm}0.7}$ &${27.9}_{{\pm}0.6}$ &${33.2}_{{\pm}1.2}$ &${33.7}_{{\pm}1.1}$ 
        
        &\XSolidBrush &${36.6}_{{\pm}0.9}$ 
        
        &${28.6}_{{\pm}1.3}$ &${32.3}_{{\pm}0.8}$ &${34.9}_{{\pm}0.9}$ 
        
        &$\mathbf{{43.2}_{{\pm}1.3}}$\\

        MSD07 &QUBIQ &${28.7}_{{\pm}1.8}$ &${30.6}_{{\pm}0.6}$ &${25.9}_{{\pm}0.8}$ &${34.5}_{{\pm}0.9}$ &${34.3}_{{\pm}1.3}$ 
        
        &\XSolidBrush &${36.5}_{{\pm}1.6}$ 
        
        &${29.3}_{{\pm}1.0}$ &${34.5}_{{\pm}1.6}$ &${35.8}_{{\pm}0.8}$ 
        
        &$\mathbf{{42.1}_{{\pm}0.6}}$\\

        KiTS21 &KiTS23 &${57.8}_{{\pm}0.6}$ &${63.0}_{{\pm}0.7}$ &${54.4}_{{\pm}0.5}$ &${66.2}_{{\pm}0.9}$ &${66.7}_{{\pm}0.6}$
        
        &\XSolidBrush &${68.4}_{{\pm}0.4}$ 
        
        &${65.8}_{{\pm}1.1}$ &${67.2}_{{\pm}0.6}$ &${68.6}_{{\pm}0.6}$ 
        
        &$\mathbf{{73.2}_{{\pm}0.8}}$\\
        
        MSD06 &RIDER &${37.5}_{{\pm}1.0}$ &${40.6}_{{\pm}0.9}$ &${36.8}_{{\pm}0.6}$ &${42.2}_{{\pm}1.2}$ &${43.8}_{{\pm}1.1}$ 

        &\XSolidBrush &\XSolidBrush 
        
        &${35.4}_{{\pm}1.0}$ &${43.9}_{{\pm}0.8}$ &${46.5}_{{\pm}1.2}$ 
        
        &$\mathbf{{55.1}_{{\pm}1.0}}$\\
        
        \toprule[1.2pt]
	\end{tabular}
    \end{threeparttable}        
\label{table_out_of_domain}
\end{sidewaystable*}

\clearpage

\begin{table*}
	\setlength{\abovecaptionskip}{0.pt}
	\setlength{\belowcaptionskip}{-0.em}
	\centering
	\footnotesize
 \caption{The number of tumors/lesions across different sizes in 12 public annotated tumor datasets.}
\begin{threeparttable}
	\begin{tabular}{l|ccccc}
		\toprule[1.2pt]
		\textbf{Diameter} &\textbf{Liver tumor} &\textbf{Pancreas tumor} &\textbf{Kidney tumor} &\textbf{Lung tumor} &\textbf{COVID-19}
        \\
        \hline
        $d<2cm$ &147 &801 &148 &89 &318\\
        
        $d~{\geq}~2cm$ &505 &3,761 &868 &206 &1,636\\
        
        \toprule[1.2pt]
	\end{tabular}
    \end{threeparttable}        
\label{table_tumor_number}
\end{table*}

\begin{table*}
	\setlength{\abovecaptionskip}{0.pt}
	\setlength{\belowcaptionskip}{-0.em}
	\centering
	\footnotesize
 \caption{Quantitative results of detecting small tumors/lesions (diameter $<$ 2cm). We report the sensitivity results for five types of tumors/lesions. The best results are \textbf{bolded} while the second best results are \underline{underlined}. The standard deviations are obtained from 5-fold evaluation.}
\begin{threeparttable}
	\begin{tabular}{lccccc}
		\toprule[1.2pt]
		\textbf{Method} &\textbf{Liver tumor} &\textbf{Pancreas tumor} &\textbf{Kidney tumor} &\textbf{Lung tumor} &\textbf{COVID-19}
        \\
        \hline
        UNet\cite{UNET} &${43.3}_{{\pm}9.0}$ &${38.3}_{{\pm}8.5}$ &${42.4}_{{\pm}8.4}$ &${40.0}_{{\pm}10.1}$ &${42.3}_{{\pm}9.3}$\\
        
        TransUNet\cite{transunet} &${40.2}_{{\pm}9.9}$ &${48.1}_{{\pm}10.8}$ &${47.4}_{{\pm}8.0}$ &${41.6}_{{\pm}7.5}$ &${45.0}_{{\pm}7.5}$\\
        
        UNETR\cite{unetr} &${38.6}_{{\pm}10.5}$ &${37.2}_{{\pm}10.3}$ &${34.1}_{{\pm}9.8}$ &${37.8}_{{\pm}10.2}$ &${35.3}_{{\pm}7.6}$\\
        
        nnUNet\cite{nnUNet} &\underline{${48.6}_{{\pm}10.7}$} &${52.5}_{{\pm}7.1}$ &${51.9}_{{\pm}10.2}$ &${48.9}_{{\pm}8.8}$ &${43.8}_{{\pm}6.0}$\\
        
        SwinUNETR\cite{swin} &${47.7}_{{\pm}11.0}$ &\underline{${53.5}_{{\pm}6.8}$} &\underline{${52.4}_{{\pm}7.9}$} &\underline{${50.5}_{{\pm}9.6}$} &\underline{${44.6}_{{\pm}6.4}$}\\
        
        \rowcolor{mygray}
        \textbf{FreeTumor} &$\mathbf{{70.6}_{{\pm}6.1}}$ &$\mathbf{{63.8}_{{\pm}6.1}}$ &$\mathbf{{69.1}_{{\pm}6.9}}$ &$\mathbf{{68.3}_{{\pm}7.7}}$ &$\mathbf{{58.6}_{{\pm}8.5}}$ \\
        
        \toprule[1.2pt]
	\end{tabular}
    \end{threeparttable}        
\label{table_early}
\end{table*}

\begin{figure}
	\centering
	\includegraphics[width=0.9\linewidth]{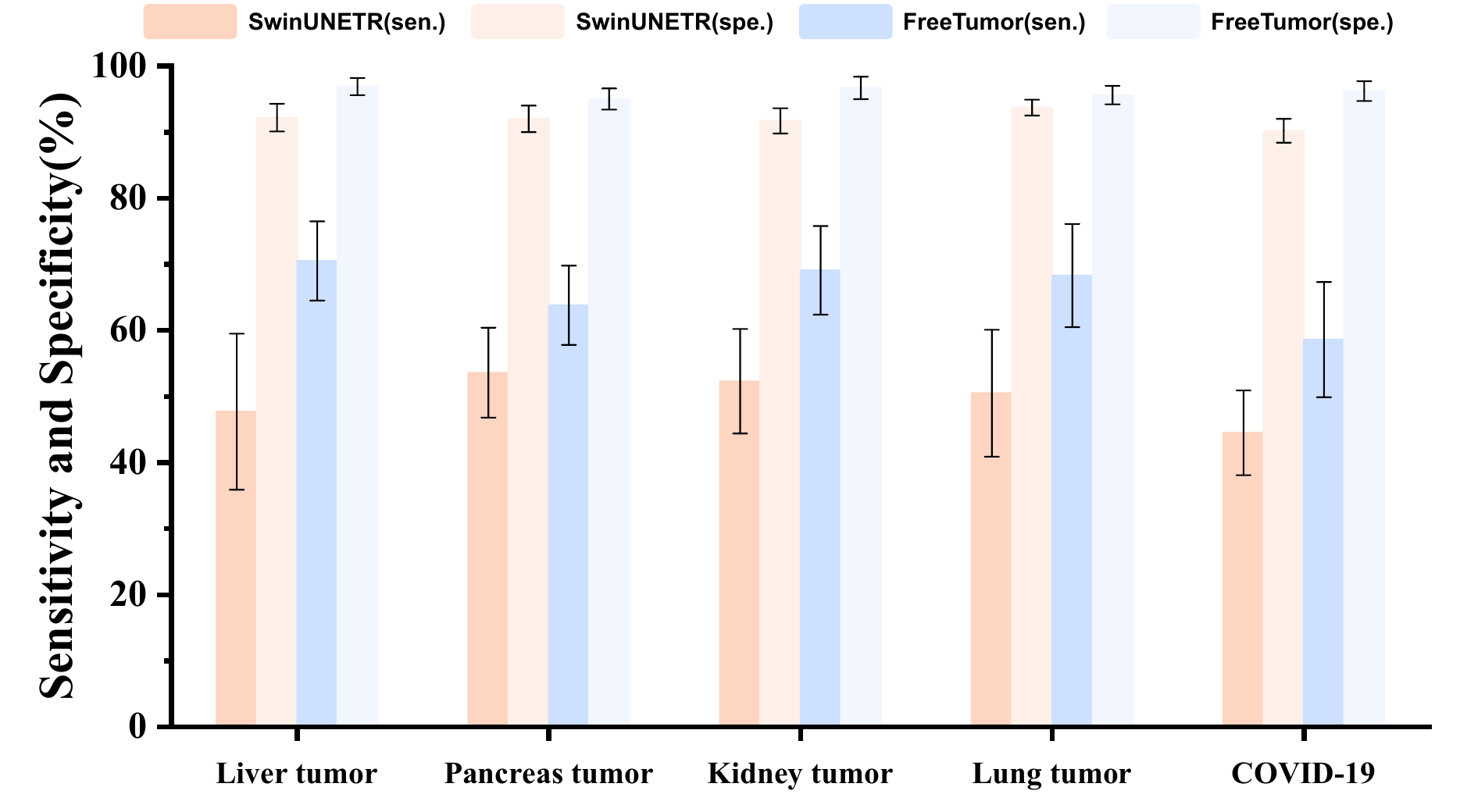}
 \vspace{-.1in}
	\caption{The sensitivity (sen.) and specificity (spe.) results of detecting small tumors/lesions (diameter $<$ 2cm). We compare with the baseline method SwinUNETR~\cite{swin} across five types of tumors/lesions. Since the small tumors (diameter $<$ 2cm) are rare in public datasets, the false alarm is relatively low and the specificity results are higher. Thus, we highlight the sensitivity results in the main paper.}
	\label{fig_early_sen_spe}
\end{figure}

\clearpage


\begin{figure}
	\centering
	\includegraphics[width=1\linewidth]{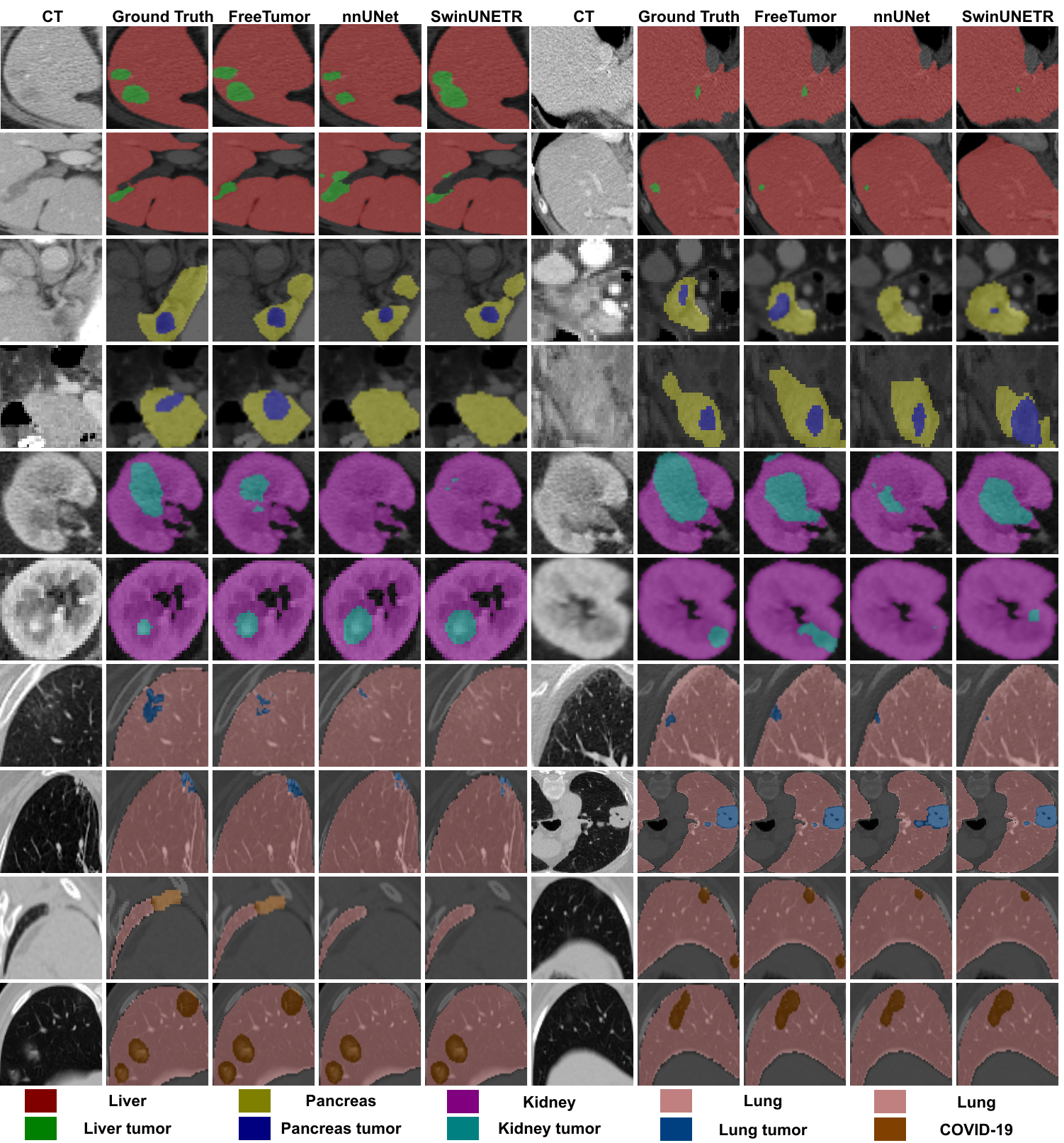}
 \vspace{-.1in}
	\caption{\textbf{Qualitative tumor segmentation results with comparison methods}. The segmentation results of nnUNet and SwinUNETR are presented for comparison. We also visualize the corresponding organ segmentation results for better understanding. The tumor regions are cropped and magnified for better visualization.}
	\label{fig_comparison}
\end{figure}
\clearpage


\renewcommand{\arraystretch}{1.4}
\begin{table*}
	\setlength{\abovecaptionskip}{0.pt}
	\setlength{\belowcaptionskip}{-0.em}
	\centering
	\footnotesize
 \caption{We investigate the data scaling law in tumor recognition, the results are shown in Figure~\ref{fig_segmentation}~(d). Specifically, for abdomen CT data, we augment the data scale from 0.7K to 20K. For chest CT data, we augment the data scale from 6K to 140K. The details of datasets utilization are shown as below.}
\begin{threeparttable}
	\begin{tabular}{cccccccc}
		\toprule[1.2pt]
		\multirow{2}{*}{\textbf{Dataset}} &\multirow{2}{*}{\textbf{Region}} &\multicolumn{4}{c}{\textbf{Data Scale}} &\multirow{2}{*}{\textbf{Number of volumes}}\\
        \cline{3-6}
        & &\textbf{0.7K} &\textbf{8K} &\textbf{13K} &\textbf{20K} &\\
		\hline
        LiTS &Abdomen &\CheckmarkBold &\CheckmarkBold &\CheckmarkBold &\CheckmarkBold &131\\
        HCC-TACE &Abdomen &\CheckmarkBold &\CheckmarkBold &\CheckmarkBold &\CheckmarkBold &104\\
        IRCAD &Abdomen &\CheckmarkBold &\CheckmarkBold &\CheckmarkBold &\CheckmarkBold &22\\
        MSD07 &Abdomen &\CheckmarkBold &\CheckmarkBold &\CheckmarkBold &\CheckmarkBold &281\\
        QUBIQ &Abdomen &\CheckmarkBold &\CheckmarkBold &\CheckmarkBold &\CheckmarkBold &40\\
        KiTS23 &Abdomen &\CheckmarkBold &\CheckmarkBold &\CheckmarkBold &\CheckmarkBold &489\\
        KIPA &Abdomen &\CheckmarkBold &\CheckmarkBold &\CheckmarkBold &\CheckmarkBold &70\\
        BTCV &Abdomen & &\CheckmarkBold &\CheckmarkBold &\CheckmarkBold &30\\
        AMOS22 &Abdomen & &\CheckmarkBold &\CheckmarkBold &\CheckmarkBold &300\\
        FLARE22 &Abdomen & &\CheckmarkBold &\CheckmarkBold &\CheckmarkBold &50\\
        WORD &Abdomen & &\CheckmarkBold &\CheckmarkBold &\CheckmarkBold &120\\
        MSD09 &Abdomen & &\CheckmarkBold &\CheckmarkBold &\CheckmarkBold &41\\
        MSD10 &Abdomen & &\CheckmarkBold &\CheckmarkBold &\CheckmarkBold &126\\
        CHAOS &Abdomen & &\CheckmarkBold &\CheckmarkBold &\CheckmarkBold &40\\
        TCIA-PANC &Abdomen & &\CheckmarkBold &\CheckmarkBold &\CheckmarkBold &80\\
        PANORAMA &Abdomen & &\CheckmarkBold &\CheckmarkBold &\CheckmarkBold &2,238\\
        FLARE23 &Abdomen & &\CheckmarkBold &\CheckmarkBold &\CheckmarkBold &4,500\\
        AbdomenAtlas &Abdomen & & &\CheckmarkBold &\CheckmarkBold &5,195\\
        Abdomen-1k &Abdomen & & & &\CheckmarkBold &1,062\\
        TotalSegmentator &Abdomen & & & &\CheckmarkBold &534\\
        DeepLesion &Abdomen & & & &\CheckmarkBold &1,618\\
        COLONOGRAPHY &Abdomen & & & &\CheckmarkBold &1,730\\
        MELA &Abdomen & & & &\CheckmarkBold &770\\
        \hline
        Total &Abdomen & & & & &19,571\\
        \toprule[1.2pt]
	\end{tabular}
    \end{threeparttable}        
\label{table_scale_abdomen_data}
\end{table*}

\begin{table*}
	\setlength{\abovecaptionskip}{0.pt}
	\setlength{\belowcaptionskip}{-0.em}
	\centering
	\footnotesize
\begin{threeparttable}
	\begin{tabular}{ccccccc}
		\toprule[1.2pt]
		\multirow{2}{*}{\textbf{Dataset}} &\multirow{2}{*}{\textbf{Region}} &\multicolumn{3}{c}{\textbf{Data Scale}} &\multirow{2}{*}{\textbf{Number of volumes}}\\
        \cline{3-5}
        & &\textbf{6K} &\textbf{56K} &\textbf{140K} &\\
		\hline
        MSD06 &Chest &\CheckmarkBold &\CheckmarkBold &\CheckmarkBold &63\\
        RIDER &Chest &\CheckmarkBold &\CheckmarkBold &\CheckmarkBold &56\\
        CV19-20 &Chest &\CheckmarkBold &\CheckmarkBold &\CheckmarkBold &199\\
        LIDC &Chest &\CheckmarkBold &\CheckmarkBold &\CheckmarkBold &589\\
        MELA &Chest &\CheckmarkBold &\CheckmarkBold &\CheckmarkBold &770\\
        LUNA16 &Chest &\CheckmarkBold &\CheckmarkBold &\CheckmarkBold &843\\
        STOIC &Chest &\CheckmarkBold &\CheckmarkBold &\CheckmarkBold &2,000\\
        StonyBrook &Chest &\CheckmarkBold &\CheckmarkBold &\CheckmarkBold &2,316\\
        CT-RATE &Chest & &\CheckmarkBold &\CheckmarkBold &50,118\\
        NLST &Chest & & &\CheckmarkBold &84,830\\
        \hline
        Total &Chest & & & &141,784\\
        
        \toprule[1.2pt]
	\end{tabular}
    \end{threeparttable}        
\label{table_scale_chest_data}
\end{table*}

\clearpage

\renewcommand{\arraystretch}{1.4}
\begin{table*}
    \setlength{\abovecaptionskip}{0pt}
    \setlength{\belowcaptionskip}{-0em}
    \centering
    \footnotesize
    \caption{\textbf{The effectiveness of data scaling law in abdomen region tumor segmentation}. We report the Dice score ($\%$) results of 3 datasets, \emph{i.e.}, LiTS, MSD07, and KiTS23. The standard deviations are obtained from five times of experiments.}
    \begin{threeparttable}
        \begin{tabular}{cccccc}
            \toprule[1.2pt]
            \multirow{2}{*}{\textbf{Dataset}} & \multirow{2}{*}{\textbf{Type}} & \multicolumn{4}{c}{\textbf{Data Scale}} \\
            \cline{3-6}
            & & \textbf{0.7K} & \textbf{8K} & \textbf{13K} & \textbf{20K} \\
            \hline
            LiTS & Liver tumor & $63.5_{{\pm}0.5}$ & $65.0_{{\pm}0.5}$ & $65.2_{{\pm}0.6}$ & $\mathbf{65.5_{{\pm}0.6}}$ \\
            MSD07 & Pancreas tumor & $55.2_{{\pm}0.6}$ & $57.8_{{\pm}0.4}$ & $58.1_{{\pm}0.3}$ & $\mathbf{58.6_{{\pm}0.5}}$ \\
            KiTS23 & Kidney tumor & $73.2_{{\pm}0.7}$ & $75.0_{{\pm}0.3}$ & $75.1_{{\pm}0.3}$ & $\mathbf{75.3_{{\pm}0.6}}$ \\
            \toprule[1.2pt]
        \end{tabular}
    \end{threeparttable}        
    \label{table_seg_scale_abdomen_data}
\end{table*}

\begin{table*}
    \setlength{\abovecaptionskip}{0pt}
    \setlength{\belowcaptionskip}{-0em}
    \centering
    \footnotesize
    \caption{\textbf{The effectiveness of data scaling law in chest region tumor segmentation}. We report the Dice score ($\%$) results of 2 datasets, \emph{i.e.}, MSD06 and CV19-20. The standard deviations are obtained from five times of experiments.}
    \begin{threeparttable}
        \begin{tabular}{ccccc}
            \toprule[1.2pt]
            \multirow{2}{*}{\textbf{Dataset}} & \multirow{2}{*}{\textbf{Type}} & \multicolumn{3}{c}{\textbf{Data Scale}} \\
            \cline{3-5}
            & & \textbf{6K} & \textbf{56K} & \textbf{140K} \\
            \hline
            MSD06 & Lung tumor & $64.3_{{\pm}1.0}$ & $65.3_{{\pm}0.7}$ & $\mathbf{65.8_{{\pm}0.7}}$ \\
            CV19-20 & COVID-19 & $60.9_{{\pm}0.6}$ & $70.5_{{\pm}0.4}$ & $\mathbf{71.2_{{\pm}0.6}}$ \\
            \toprule[1.2pt]
        \end{tabular}
    \end{threeparttable}        
    \label{table_seg_scale_chest_data}
\end{table*}


\begin{table*}
	\setlength{\abovecaptionskip}{0.pt}
	\setlength{\belowcaptionskip}{-0.em}
	\centering
	\footnotesize
 \caption{\textbf{Previous synthesis methods fail to effectively leverage large-scale data for segmentation training}. 
 We employ these two synthesis models to synthesize tumors on our new datasets for segmentation training. However, without large-scale synthesis training, these methods fail to generalize well on unseen datasets with different image characteristics.
 We report the Dice scores ($\%$) on LiTS~\cite{LITS}, MSD07~\cite{msd}, and KiTS23~\cite{kits} datasets. 20K is the total number of abdomen CT volumes used for training liver, pancreas, and kidney tumor models. More quantitative and qualitative comparisons are presented in Table~\ref{table_compare_synthesis} and Figure~\ref{fig_comparison}.}
\begin{threeparttable}
	 \begin{tabular}{l|cc|ccc}
        \toprule[1.2pt]
		\textbf{Method} &\textbf{Syn. Training Scale} &\textbf{Seg. Training Scale} &\textbf{LiTS} &\textbf{MSD07} &\textbf{KiTS23}\\
        \hline
        SynTumor &\XSolidBrush &0.1K &$60.2_{{\pm}1.1}$ &\XSolidBrush &\XSolidBrush\\
        SynTumor &\XSolidBrush &20K &$52.8_{{\pm}2.3}$(\textcolor{red}{$\downarrow$}) &\XSolidBrush &\XSolidBrush\\
        \hline
        DiffTumor &360 &360 &$62.3_{{\pm}0.7}$ &$54.5_{{\pm}1.8}$ &$70.4_{{\pm}0.9}$\\
        DiffTumor &360 &20K &$60.8_{{\pm}1.2}$(\textcolor{red}{$\downarrow$}) &$38.6_{{\pm}2.1}$(\textcolor{red}{$\downarrow$}) &$69.2_{{\pm}1.6}$(\textcolor{red}{$\downarrow$})\\
        \hline
        \rowcolor{mygray}
        FreeTumor &20K &20K &$\mathbf{65.5_{{\pm}0.6}}$ &$\mathbf{58.6_{{\pm}1.2}}$ &$\mathbf{75.3_{{\pm}0.8}}$\\
        \toprule[1.2pt]
        \end{tabular}
    \end{threeparttable}        
\label{table_abla_scale_other_synthesis}
\end{table*}

\clearpage

\begin{table*}
	\setlength{\abovecaptionskip}{0.pt}
	\setlength{\belowcaptionskip}{-0.em}
	\centering
	\footnotesize
 \caption{\textbf{Ablation studies of the threshold $T$ in quality control (Section~\ref{sec_method_filter})}. SwinUNETR~\cite{swin} is the baseline model without tumor synthesis for segmentation training. We report the Dice score ($\%$) results of the LiTS dataset. The standard deviations are obtained from five times of experiments.}
\begin{threeparttable}
	 \begin{tabular}{cccc}
        \toprule[1.2pt]
		\textbf{Method} &\textbf{Synthetic} &\textbf{Quality control} &\textbf{Dice scores($\%$)}\\
        \hline
        SwinUNETR &\XSolidBrush &\XSolidBrush &$58.6_{{\pm}0.5}$\\
        \hline
        SynTumor &\CheckmarkBold &\XSolidBrush &$60.2_{{\pm}1.1}$\\
        DiffTumor &\CheckmarkBold &\XSolidBrush &$62.3_{{\pm}0.7}$\\
        \hline
        \multirow{4}{*}{FreeTumor} &\CheckmarkBold &\XSolidBrush &$62.3_{{\pm}0.4}$\\
        &\CheckmarkBold &{$T=0.5$} &$63.5_{{\pm}0.5}$\\
        &\CheckmarkBold &{$T=0.7$} &$\mathbf{65.5_{{\pm}0.6}}$\\
        &\CheckmarkBold &{$T=0.9$} &$64.0_{{\pm}0.3}$\\
        \toprule[1.2pt]
        \end{tabular}
    \end{threeparttable}        
\label{table_abla_filtering}
\end{table*}

\begin{table*}
	\setlength{\abovecaptionskip}{0.pt}
	\setlength{\belowcaptionskip}{-0.em}
	\centering
	\footnotesize
 \caption{\textbf{Ablation studies of loss functions}. We report the Dice score ($\%$) results of the LiTS dataset. The standard deviations are obtained from five times of experiments.}
\begin{threeparttable}
	 \begin{tabular}{cccc}
    \toprule[1.2pt]
		\multicolumn{2}{c}{\textbf{Loss}} &\multirow{2}{*}{\textbf{Filtering}} &\multirow{2}{*}{\textbf{Dice scores($\%$)}}\\
        \cline{1-2}
        {$L_{seg}$} &{$L_{cls}$} &\\
		\hline
        \XSolidBrush &\XSolidBrush &\XSolidBrush &$58.6_{{\pm}0.5}$\\
        \hline
        \CheckmarkBold &\XSolidBrush &\XSolidBrush &$62.0_{{\pm}0.3}$\\
        \XSolidBrush &\CheckmarkBold &\XSolidBrush &$61.8_{{\pm}0.3}$\\
        \CheckmarkBold &\CheckmarkBold &\XSolidBrush &$64.3_{{\pm}0.5}$\\
        \CheckmarkBold &\CheckmarkBold &\CheckmarkBold &$\mathbf{65.5_{{\pm}0.6}}$\\
        \toprule[1.2pt]
    \end{tabular}
    \end{threeparttable}        
\label{table_abla_adversarial}
\end{table*}

\clearpage
\newpage

\renewcommand{\arraystretch}{1.5}
\begin{table*}
	\setlength{\abovecaptionskip}{0.pt}
	\setlength{\belowcaptionskip}{-0.em}
	\centering
	\footnotesize
 \caption{We collect 161,310 publicly available CT volumes from 33 different sources to train our FreeTumor model. These datasets mainly cover abdomen and chest regions, while only $2.3\%$ of them contain annotated tumors. Specifically, the abdomen datasets are used for liver, pancreas, and kidney tumors, while the chest datasets are used for lung tumors and COVID-19. The descriptions of these datasets are provided in the original links.}
\begin{threeparttable}
	\begin{tabular}{llll}
		\toprule[1.2pt]
		\textbf{Dataset} &\textbf{Description} &\textbf{Num.} &\textbf{Link}\\
        \hline
        \textbf{With annotated tumors}\\
        LiTS\cite{LITS} &Liver Tumor &131 &\href{https://competitions.codalab.org/competitions/17094}{competitions.codalab.org/competitions/17094}\\
        HCC-TACE\cite{HCC-TACE-Seg} &Liver Tumor &104 &\href{https://www.cancerimagingarchive.net/collection/hcc-tace-seg}{cancerimagingarchive.net/collection/hcc-tace}\\
        IRCAD\cite{3D-IRCADb} &Liver Tumor &22 &\href{https://www.ircad.fr/research/data-sets/liver-segmentation-3d-ircadb-01}{www.ircad.fr/research/data-sets}\\
        MSD07-Pancreas\cite{msd} &Pancreas Tumor &281 &\href{https://decathlon-10.grand-challenge.org/}{decathlon-10.grand-challenge.org}\\
        PANORAMA\cite{PANORAMA} &Pancreas Tumor &2,238 &\href{https://panorama.grand-challenge.org/}{panorama.grand-challenge.org}\\
        QUBIQ\cite{Qubiq} &Pancreas Tumor &40 &\href{https://qubiq21.grand-challenge.org/QUBIQ/}{qubiq21.grand-challenge.org/QUBIQ}\\
        KiTS23\cite{kits} &Kidney Tumor &489 &\href{https://kits-challenge.org/kits23/}{kits-challenge.org/kits23}\\
        KIPA\cite{KIPA} &Kidney Tumor &70 &\href{https://kipa22.grand-challenge.org/}{kipa22.grand-challenge.org}\\
        MSD06-Lung\cite{msd} &Lung Tumor &63 &\href{https://decathlon-10.grand-challenge.org/}{decathlon-10.grand-challenge.org}\\
        RIDER\cite{RIDER-LungCT-Seg} &Lung Tumor &59 &\href{https://www.cancerimagingarchive.net/analysis-result/rider-lungct-seg/}{cancerimagingarchive.net/analysis-result/rider}\\
        CV19-20\cite{COVID} &COVID-19 &199 &\href{https://covid-segmentation.grand-challenge.org/}{covid-segmentation.grand-challenge.org}\\
        
        \hline
        \textbf{Without annotated tumors} & & &\\
        BTCV\cite{btcv} &Abdomen &30 &\href{https://www.synapse.org/#!Synapse:syn3193805/wiki/}{synapse.org/\#!Synapse:syn3193805/wiki}\\
        AMOS22\cite{amos} &Abdomen &300 &\href{https://amos22.grand-challenge.org/}{amos22.grand-challenge.org}\\
        FLARE22\cite{FLARE} &Abdomen &50 &\href{https://flare22.grand-challenge.org/}{flare22.grand-challenge.org}\\
        WORD\cite{word} &Abdomen &120 &\href{https://github.com/HiLab-git/WORD}{github.com/HiLab-git/WORD}\\
        MSD09-Spleen\cite{msd} &Abdomen &41 &\href{https://decathlon-10.grand-challenge.org/}{decathlon-10.grand-challenge.org}\\
        MSD10-Colon\cite{msd} &Abdomen &126 &\href{https://decathlon-10.grand-challenge.org/}{decathlon-10.grand-challenge.org}\\
        CHAOS~\cite{CHAOS} &Abdomen &40 &\href{https://chaos.grand-challenge.org/Combined_Healthy_Abdominal_Organ_Segmentation/}{chaos.grand-challenge.org}\\
        TCIA-Panc\cite{tcia-panc} &Abdomen &80 &\href{https://www.cancerimagingarchive.net/collection/pancreas-ct/}{cancerimagingarchive.net/collection/pancreas}\\
        Abdomenct-1k\cite{AbdomenCT-1K} &Abdomen &1,062 &\href{https://github.com/JunMa11/AbdomenCT-1K}{github.com/JunMa11/AbdomenCT-1K}\\
        TotalSegmentator\cite{total} &Abdomen &534 &\href{https://github.com/wasserth/TotalSegmentator}{github.com/wasserth/TotalSegmentator} \\
        DeepLesion~\cite{DeepLesion} &Abdomen &1,618 &\href{https://uls23.grand-challenge.org/}{uls23.grand-challenge.org}\\
        COLONOGRAPHY\cite{CT-COLONOGRAPHY} &Abdomen/Chest &1,730 &\href{https://doi.org/10.7937/K9/TCIA.2015.NWTESAY1}{doi.org/10.7937/K9/TCIA.2015.NWTESAY1}\\
        FLARE23\cite{FLARE} &Abdomen &4,500 &\href{https://codalab.lisn.upsaclay.fr/competitions/12239}{codalab.lisn.upsaclay.fr/competitions/12239}\\
        AbdomenAtlas\cite{atlas,bassi2025touchstone,wanglarge} &Abdomen &5,195 &\href{https://github.com/MrGiovanni/AbdomenAtlas}{github.com/MrGiovanni/AbdomenAtlas}\\
        MELA\cite{MELA} &Abdomen/Chest &770 &\href{https://doi.org/10.5281/zenodo.6575197}{doi.org/10.5281/zenodo.6575197}\\
        LIDC\cite{LIDC} &Chest &589 &\href{https://www.cancerimagingarchive.net/collection/lidc-idri/}{cancerimagingarchive.net/collection/lidc-idri}\\
        TCIA-Covid\cite{tcia} &Chest &722 &\href{https://www.cancerimagingarchive.net/collection/covid-19-ar/}{cancerimagingarchive.net/collection/covid-19}\\
        LUNA16\cite{LUNA} &Chest &843 &\href{https://luna16.grand-challenge.org/Home/}{luna16.grand-challenge.org/Home}\\
        STOIC 2021\cite{STOIC} &Chest &2,000 &\href{https://stoic2021.grand-challenge.org/}{stoic2021.grand-challenge.org}\\
        StonyBrook\cite{StonyBrookChestCT} &Chest &2,316 &\href{https://doi.org/10.7937/TCIA.BBAG-2923}{doi.org/10.7937/TCIA.BBAG-2923}\\
        CT-RATE\cite{ctrate} &Chest &50,118 &\href{https://huggingface.co/datasets/ibrahimhamamci/CT-RATE}{huggingface.co/datasets/CT-RATE}\\
        NLST\cite{NLST} &Chest &84,830 &\href{https://doi.org/10.7937/TCIA.HMQ8-J677}{doi.org/10.7937/TCIA.HMQ8-J677}\\
        \hline
        \textbf{Total} & &\textbf{161,310} &\\
        
        \toprule[1.2pt]
	\end{tabular}
    \end{threeparttable}        
\label{table_dataset}
\end{table*}

\clearpage
\newpage

\begin{table*}[!h]
    \centering
\caption{\textbf{The public codes of methods used in this study.} 
}
\label{table_code_source}
    \begin{tabular}{ll}
    \toprule[1.2pt]
 \textbf{Method} &\textbf{Sources}\\
 \hline
         FID & \url{https://github.com/mseitzer/pytorch-fid}\\
         MONAI & \url{https://github.com/Project-MONAI/research-contributions}\\
         nnUNet& \url{https://github.com/MIC-DKFZ/nnUNet}\\
        TransUNet&\url{https://github.com/Beckschen/TransUNet}\\
         VoCo& \url{https://github.com/Luffy03/Large-Scale-Medical}\\
         Lungmask& \url{https://github.com/JoHof/lungmask}\\
 SynTumor&\url{https://github.com/MrGiovanni/SyntheticTumors}\\
 DiffTumor&\url{https://github.com/MrGiovanni/DiffTumor}\\
 \toprule[1.2pt]
    \end{tabular}

\end{table*}

\begin{table*}[!h]
    \centering
\caption{\textbf{Pre-processing details and Training settings.} 
}
\label{table_preprocess}
    \begin{tabular}{l|l}
    \toprule[1.2pt]
 Clip Hounsfield Unit for abdomen CT &(-175, 250)\\
 Crop Size for abdomen CT &(96, 96, 96)\\
 \hline
 Clip Hounsfield Unit for chest CT &(-1000, 500)\\
 Crop Size for chest CT &(192, 192, 32)\\   
\toprule[1.2pt]
 Network architecture &SwinUNETR \\
 Network Params &72M \\
 Segmentation Loss &Dice-CE\\
 Optimizer &AdamW\\
 Batch size &4\\
 Scheduler &Cosine\\
 Learning rate (Synthesis training) &1e-4\\
 Learning rate (Segmentation training) &3e-4\\
 Training epochs (Synthesis training) &100\\
 Training epochs (Segmentation training) &100\\
\toprule[1.2pt]
    \end{tabular}
\end{table*}




\end{appendices}

\end{document}